\documentclass[aps,prd,twocolumn,twoside,superscriptaddress,floatfix,nofootinbib,natbib]{revtex4}
\usepackage{graphicx}
\usepackage{subfigure}
\usepackage{amsmath}
\usepackage{graphicx}
\usepackage{amssymb}
\usepackage{sidecap}
\usepackage{color}
\usepackage{enumerate}
\usepackage{algorithm}
\usepackage{algpseudocode}
\usepackage{verbatim}

\def\dalemb#1#2{{\vbox{\hrule height.#2pt
        \hbox{\vrule width.#2pt height#1pt \kern#1pt \vrule width.#2pt}
        \hrule height.#2pt}}}

\def\ba{\begin{eqnarray}}
\def\ea{\end{eqnarray}}
\def\be{\begin{equation}}
\def\ee{\end{equation}}

\def\gtorder{\mathrel{\raise.3ex\hbox{$>$}\mkern-14mu
             \lower0.6ex\hbox{$\sim$}}}
\def\ltorder{\mathrel{\raise.3ex\hbox{$<$}\mkern-14mu
             \lower0.6ex\hbox{$\sim$}}}

\def\be{\begin{equation}}
\def\ee{\end{equation}}

\def\gtorder{\mathrel{\raise.3ex\hbox{$>$}\mkern-14mu
             \lower0.6ex\hbox{$\sim$}}}
\def\ltorder{\mathrel{\raise.3ex\hbox{$<$}\mkern-14mu
             \lower0.6ex\hbox{$\sim$}}}

\newcommand{\mnu}  {$\Sigma m_{\nu}$}
\newcommand{\mnum}  {\Sigma m_{\nu}}

\begin{document}
\title{Towards a cosmological neutrino mass detection}
\author{R.~Allison\thanks{E-mail:~\texttt{rupert.allison@astro.ox.ac.uk}}} \address{Sub-department of Astrophysics, University of Oxford, Denys Wilkinson Building, Oxford, OX1 3RH, UK}
\author{P.~Caucal} \affiliation{Master ICFP, D\'{e}partement de Physique, \'{E}cole Normale Sup\'{e}rieure, \\24 rue Lhomond, 75005 Paris, France} \affiliation{Sub-department of Astrophysics, University of Oxford, Denys Wilkinson Building, Oxford, OX1 3RH, UK}
\author{E.~Calabrese} \affiliation{Sub-department of Astrophysics, University of Oxford, Denys Wilkinson Building, Oxford, OX1 3RH, UK}  \affiliation{Department of Astrophysical Sciences, Peyton Hall, Princeton University, Princeton, NJ USA 08544}
\author{J.~Dunkley} \affiliation{Sub-department of Astrophysics, University of Oxford, Denys Wilkinson Building, Oxford, OX1 3RH, UK}
\author{T.~Louis} \affiliation{Sub-department of Astrophysics, University of Oxford, Denys Wilkinson Building, Oxford, OX1 3RH, UK}

\begin{abstract}
Future cosmological measurements should enable the sum of neutrino masses to be determined indirectly through their effects on the expansion rate of the Universe and the clustering of matter. We consider prospects for the gravitationally lensed Cosmic Microwave Background anisotropies and Baryon Acoustic Oscillations in the galaxy distribution, examining how the projected uncertainty of $\approx15$~meV on the neutrino mass sum (a 4$\sigma$ detection of the minimal mass) might be reached over the next decade. The current 1$\sigma$ uncertainty of $\approx  103$~meV ({\sl Planck}-2015+BAO-15) will be improved by upcoming `Stage-3' CMB experiments (S3+BAO-15: 44~meV), then upcoming BAO measurements (S3+DESI: 22~meV), and planned next-generation `Stage 4' CMB experiments (S4+DESI: 15-19~meV, depending on angular range). An improved optical depth measurement is important: the projected neutrino mass uncertainty increases to $26$~meV if S4 is limited to $\ell>20$ and combined with current large-scale polarization data. Looking beyond $\Lambda$CDM, including curvature uncertainty increases the forecast mass error by $\approx$~50\% for S4+DESI, and more than doubles the error with a two-parameter dark energy equation of state. Complementary low-redshift probes including galaxy lensing will play a role in distinguishing between massive neutrinos and a departure from a $w=-1$, flat geometry.
\end{abstract}
\maketitle

\section{Introduction}
A central goal in both cosmology and particle physics is to measure the mass of the neutrino particles. The neutrino sector is still poorly understood and the mechanism that gives rise to their mass is unknown. There are thought to be three active neutrino species, with mass differences measured through solar, atmospheric, reactor and accelerator neutrino oscillation experiments (for reviews see e.g., \citet{gonzalez-garcia/nir:2003,Maltoni:2004,smirnov:2006,Feldman:2012}). The results imply a minimum total mass of $60$~meV in a normal hierarchy with two lighter and one heavier neutrino, or $100$~meV in an inverted hierarchy with two massive neutrinos.

Cosmology provides an indirect probe of massive neutrinos \citep[e.g.,][]{Hu:2002,dolgov:2002,elgaroy/lahav:2005,tegmark:2005,Hannestad:2005,Ichikawa:2005,lesgourgues/pastor:2006,fukugita:2006,Giunti:2007,abdalla/rawlings:2007,komatsu/etal:2009,Ferroni:2009,Hannestad:2010}. Massive neutrinos behave initially like non-interacting relativistic particles, and then later like cold dark matter. As such they affect the expansion rate of the Universe, compared to a pure radiation or pure matter component, as well as modifying the evolution of perturbations at early times. They also modify the growth of structure through a suppression of the clustering of matter on scales that entered the cosmic horizon while the neutrinos were relativistic. 

The current indirect 95\% upper limit from cosmological data on the sum of the neutrino masses is $\Sigma m_\nu<230$~meV from the {\sl Planck} measurements of the Cosmic Microwave Background (CMB), combined with Baryon Acoustic Oscillation (BAO) measurements from the Baryon Oscillation Spectroscopic Survey (BOSS) \citep{Anderson:2014,Planck2015XIII}. The limit is $\Sigma m_\nu<680$~meV from the CMB alone \citep{Planck2015XIII}.
Tighter limits have been found including Lyman-$\alpha$ forest measurements from quasars in the BOSS survey ($\Sigma m_\nu<120$~meV) \citep{Palanque:2013,Palanque:2015}, but the result depends on numerical hydrodynamical simulations which may contribute additional systematic uncertainty.

Recent forecasts of mass limits for upcoming cosmological datasets, including galaxy lensing and clustering, redshift-space distortions, the kinematic Sunyaev-Zel'dovich effect, and counts of galaxy clusters, have been studied extensively (e.g., \citet{Font-Ribera:2014,villaescusa-navarro/etal:2015,Mueller:2015}), showing the promise of a wide range of future cosmological data to target a neutrino mass measurement. In this paper we focus on the combination of lensed CMB and BAO measurements, datasets which do not require detailed modelling of non-linear structure formation, or an understanding of galaxy bias. The gravitationally lensed CMB measures the growth of structure at times typically before the Universe was half its current age, and on angular scales larger than $\approx100$~Mpc, so is dominated by linear physics. Studies of this combination have been reported in \citet{Hall:2012, Abazajian:2013, Wu:2014} and \citet{Pan:2015}, with a 4$\sigma$ detection of neutrino mass forecast for the next generation of experiments. In this paper we investigate this further, exploring the dependence on experimental details and on parameter degeneracies. 

 In \S\ref{sec:theory} we give a brief review of the cosmological effects of neutrinos, and in \S\ref{sec:lcdm} study how the mass measurement may be reached step-wise using data collected during the coming decade. In \S\ref{sec:depExpDetails} we investigate the dependence on experimental details, and in \S\ref{sec:unique} we explore degeneracies with other cosmological parameters. We conclude in \S\ref{sec:discussion}.

\section{Cosmological effects of neutrinos}
\label{sec:theory}

\begin{figure*}
\centering
\hskip -0.40in
\includegraphics[width=86mm]{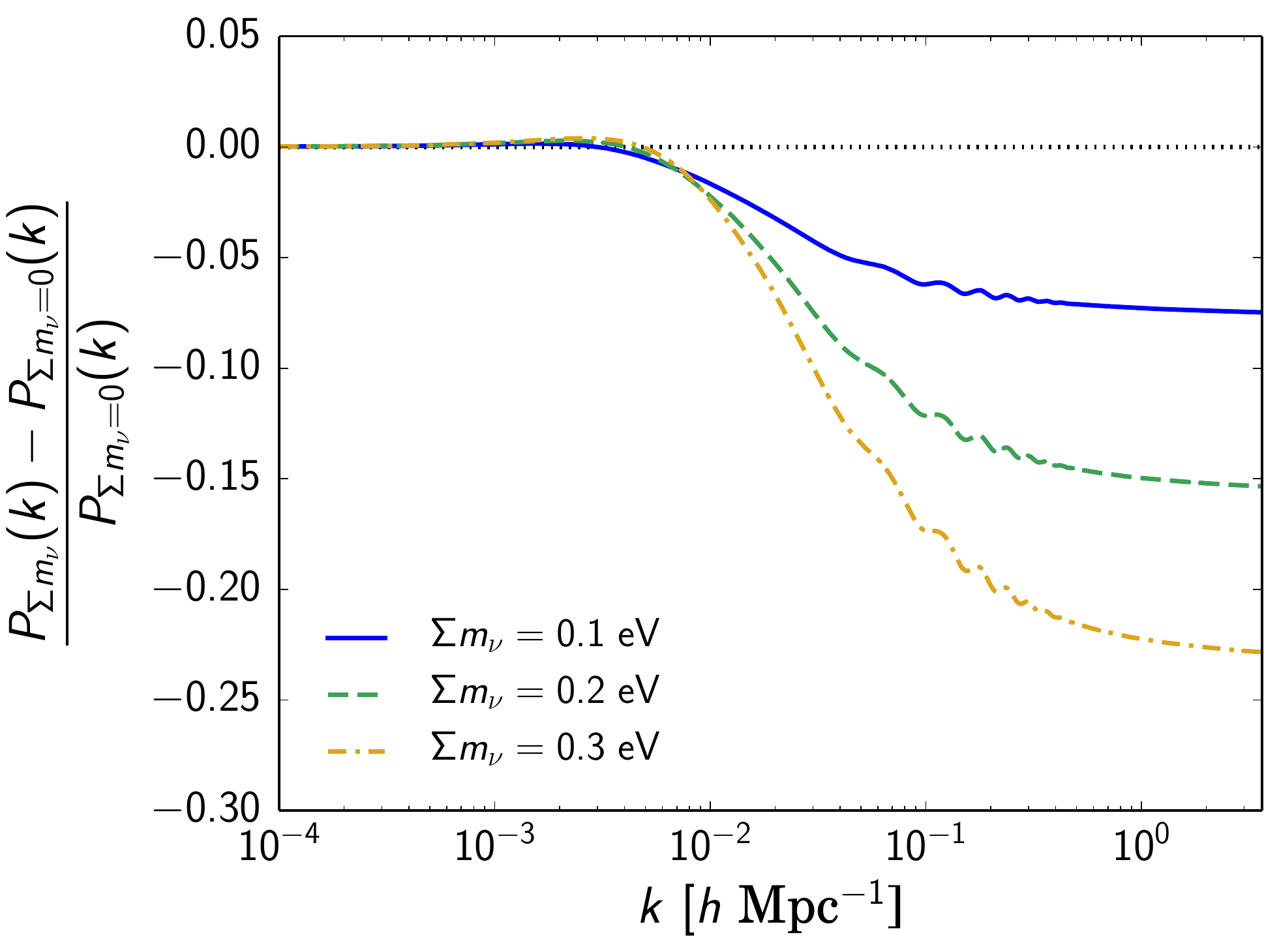}
\hskip 0.1in
\includegraphics[width=78mm]{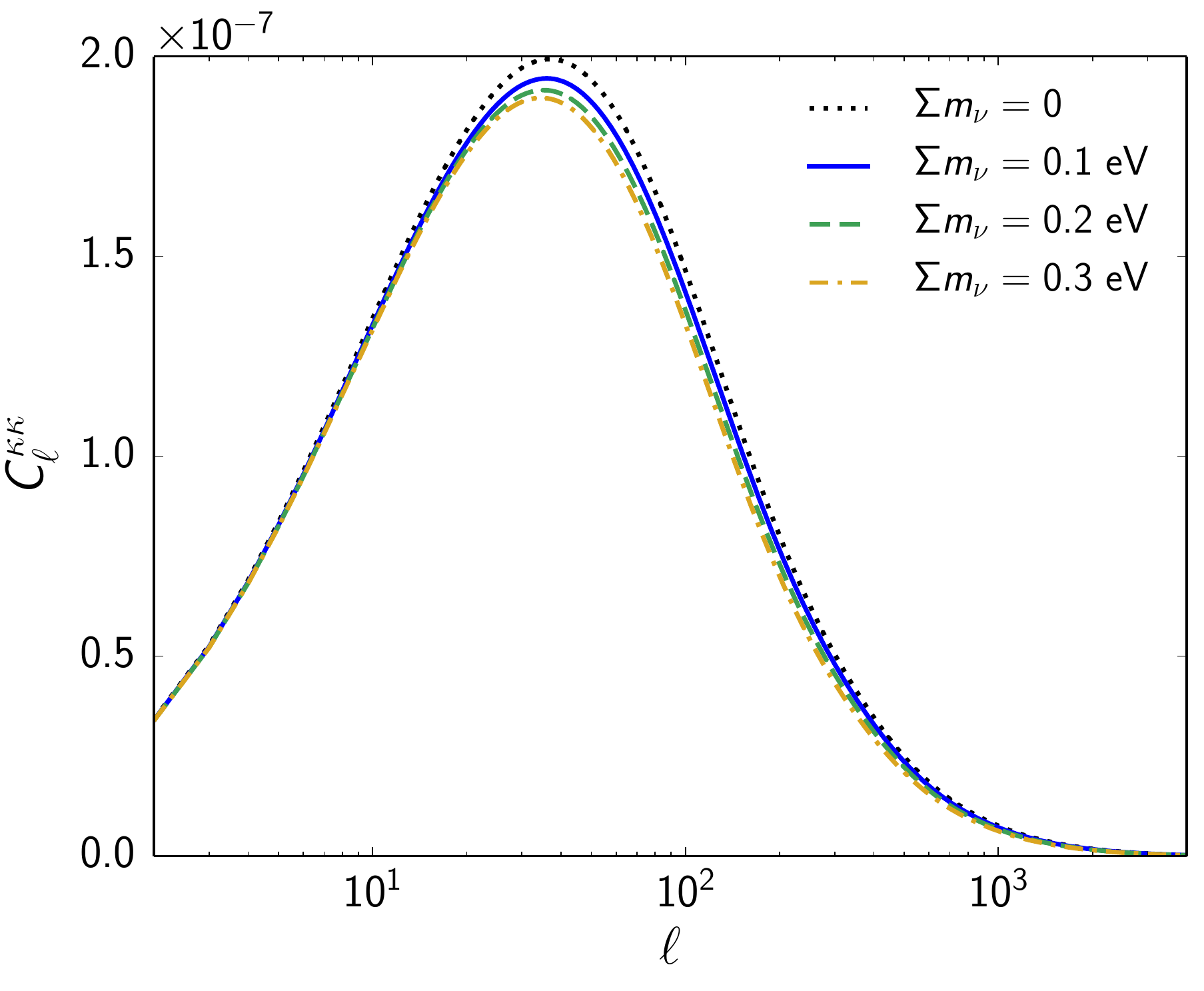}
\includegraphics[width=79mm]{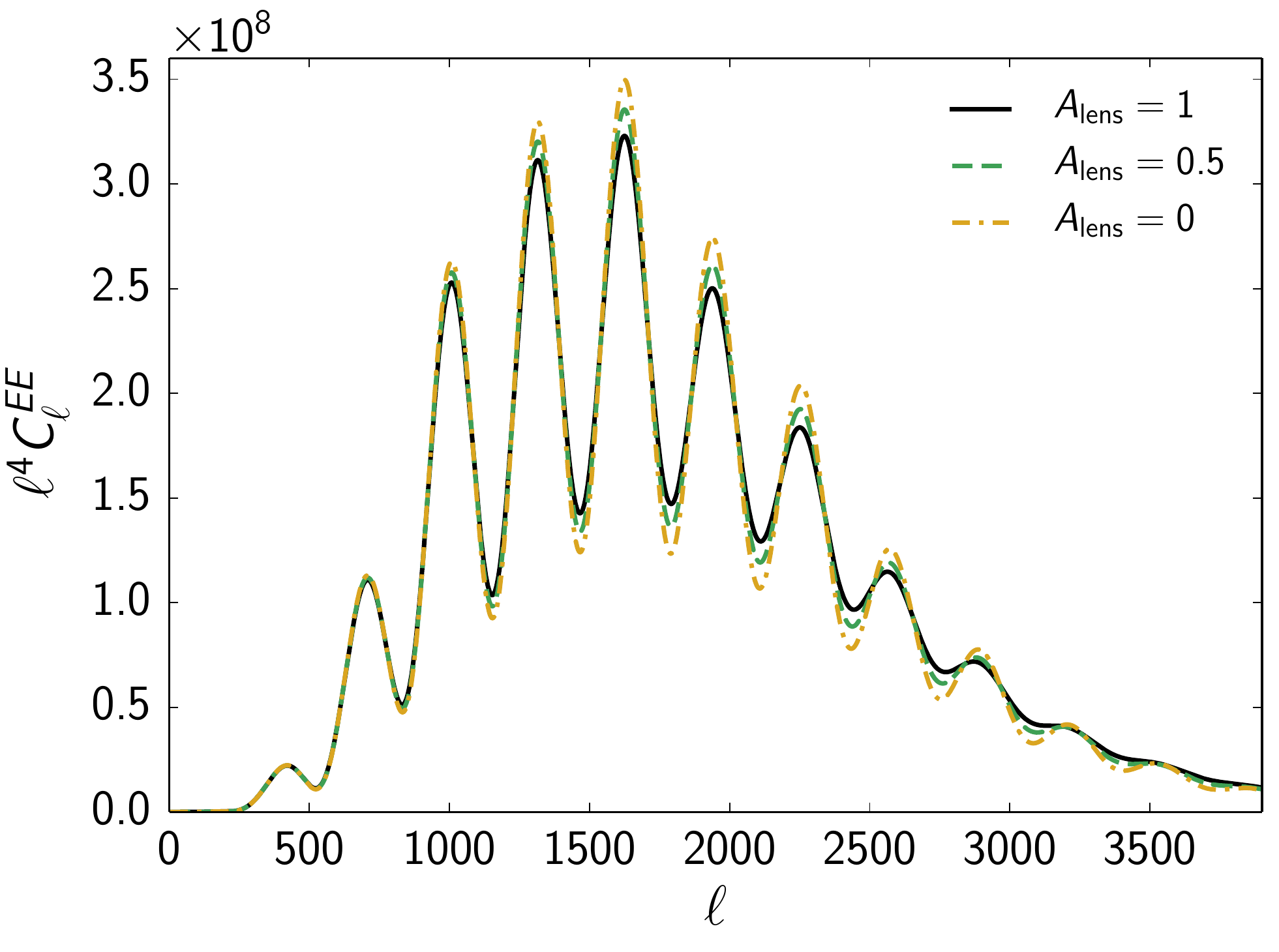}
\includegraphics[width=80mm]{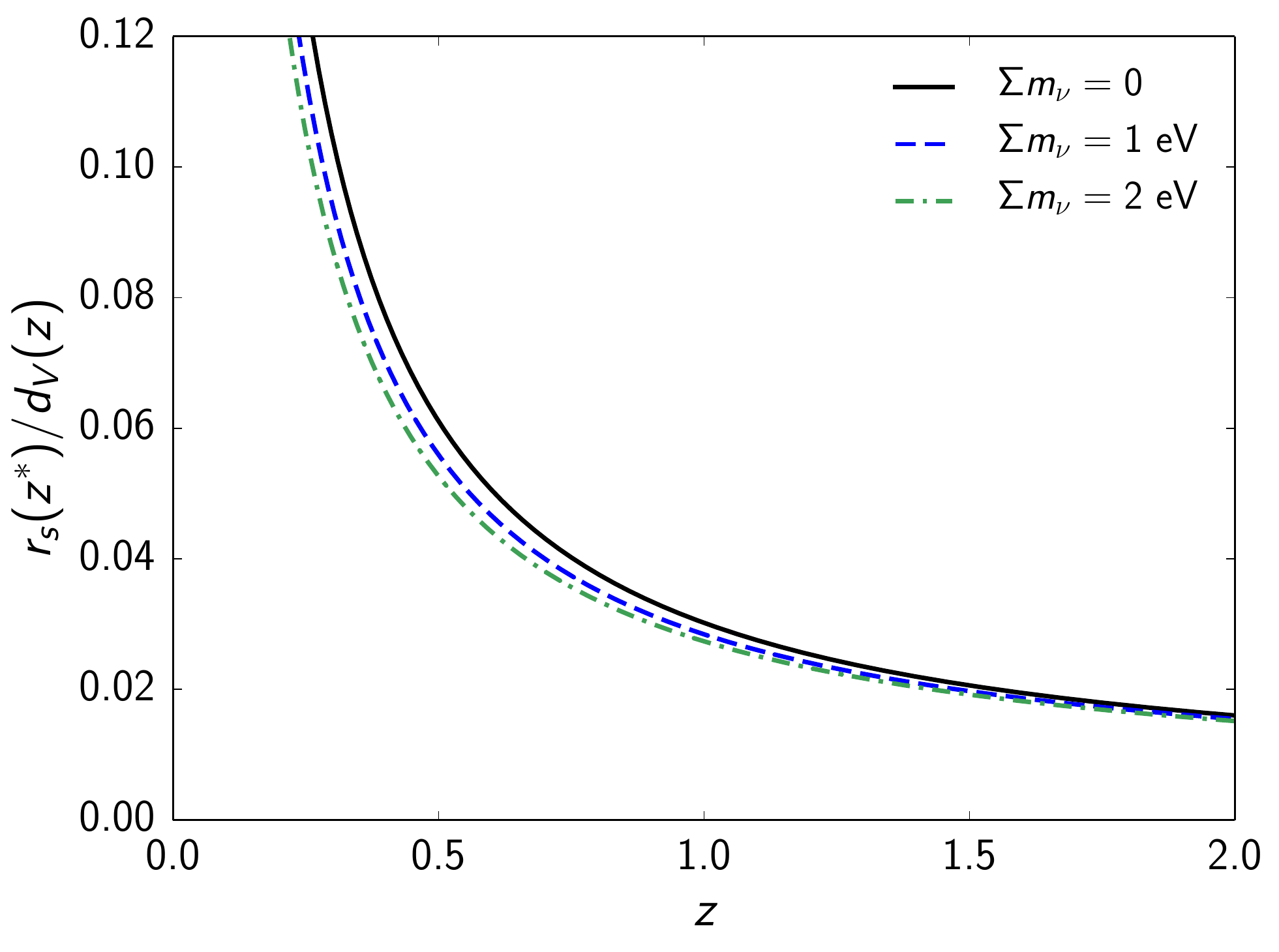}
\caption{Effect of neutrino mass on CMB power spectra and BAO distance scales. {\sl Top:} Fractional change of the matter power spectrum today $P(k)$ (left) and  CMB convergence power spectrum $C_l^{\kappa \kappa}$ (right) with neutrino mass $\Sigma m_\nu$, for fixed physical dark matter density, $\Omega_{\rm c}h^2 + \Omega_{\nu}h^2$. Suppression of power is due to neutrino free-streaming. {\sl Bottom left:} Lensed CMB $E$-mode power spectrum with varying amplitudes of the lensing potential $A_{\rm lens}$, approximating and exaggerating the effect that massive neutrinos have on the CMB polarization spectrum. {\sl Bottom right:} BAO distance ratio $r_s/d_V$ for fixed $\theta_A$ and $\Omega_ch^2$. Massive neutrinos behave like additional matter in the BAO redshift range, decreasing  $H_0$ and increasing the volume distance $d_V$. 
}
\label{figNeuMassFracDiff}
\end{figure*}

Standard Model neutrinos are initially relativistic, following a thermal distribution after decoupling from the primordial plasma when the Universe had a temperature of around $k_B T \approx 1$ MeV. The neutrino temperature decreases as the scale factor grows, until their rest-mass energy dominates and they become indistinguishable from cold dark matter. For a neutrino of mass $m_\nu$ the transition occurs at $z\approx 120(\frac{m_\nu}{60 {\rm meV}})$ \citep{Ichikawa:2005}, so current limits indicate a transition epoch of $120 \lesssim z \lesssim 460$ for a normal mass hierarchy. 

This limit implies that the neutrinos were still relativistic when the CMB decoupled, so they would be indistinguishable from massless neutrinos in the primary anisotropies. However, higher mass neutrinos become non-relativistic sooner, which reduces the early-time Integrated Sachs-Wolfe (ISW) effect. This gravitational redshift of the CMB photons arises while the non-negligible radiation component causes the potentials of the density fluctuations to evolve \citep{Hu:2002, Ichikawa:2005} and affects the anisotropies on scales around the first acoustic peak \citep{Hou:2014}. 

This effect is not sensitive to masses which remain relativistic until well after decoupling \citep{Ichikawa:2005,Komatsu:2011}, but further information comes from probes of later-time large scale structure measurements. Massive neutrinos interact weakly, allowing them to free-stream out of overdensities while relativistic, so the growth rate of matter perturbations inside the horizon is suppressed compared to a universe with only cold dark matter. For comoving wavenumbers $k \gg k_{\rm FS}$, \citet{Hu:1998} show that the suppression of the matter power spectrum today, $P(k)$, is proportional to the sum of the neutrino masses:
\begin{equation}
\label{eqSuppression}
\frac{P_{\Sigma m_\nu}(k) - P_{\Sigma m_\nu = 0}(k)}{P_{\Sigma m_\nu = 0}(k)} \approx -0.07 \left(\frac{\Sigma m_\nu}{0.1 {\rm eV}}\right) \left(\frac{\Omega_m{h^2}}{0.136}\right)^{-1},
\end{equation}
where the comoving free-streaming scale is given by 
\begin{equation}
k_{\rm FS} = 0.0072\left( \frac{\Sigma m_\nu}{0.1 {\rm eV}} \right)^{1/2} \left( \frac{\Omega_m}{ 0.315} \right)^{1/2} h {\rm \hspace{0.8mm}Mpc}^{-1},
\end{equation}
as illustrated in Fig.~\ref{figNeuMassFracDiff} and e.g., \citep{Abazajian:2013}, for models with fixed total matter density. For current limits this scale is estimated to lie in the range $0.005 \lesssim k_{\rm FS} \lesssim 0.011$~[$h$~Mpc$^{-1}$]. 

The suppression of small-scale power can be probed using galaxy clustering and the gravitational lensing of galaxies. These are promising avenues for neutrino mass measurements \citep[e.g.,][]{Font-Ribera:2014}, although these observables are sensitive to non-linearities in the matter power spectrum and scale-dependent galaxy and shape biases \cite{Mandelbaum:2015}. An alternative route is through the gravitational lensing of the CMB (see e.g., \citep{Lewis:2008} for a review). Here the CMB photons are deflected by the large-scale structure, integrated over the photon path since decoupling. 

Following \citep{Lewis:2008}, the CMB convergence angular power spectrum, $C_l^{\kappa \kappa}$, is a weighted projection of the matter power spectrum $P(k, \chi)$; under the Limber approximation, 
\begin{equation}
C_\ell^{\kappa \kappa} = \int_0^{\chi_H} d\chi \frac{W^2(\chi)}{f^2_k(\chi)}  P\left(\frac{\ell}{f_k(\chi)}, \chi \right), 
\label{eqPS}
\end{equation}
where $\chi_H$ is the comoving horizon size, $f_k(\chi)$ relates line-of-sight comoving distances and transverse comoving distances in a curved universe, 
and the window function $W(\chi)$ is 
\begin{equation}
W(\chi) = \frac{3\Omega_m H_0^2}{2c^2} \frac{f_k(\chi) f_k(\chi^* - \chi)}{a(\chi)f_k(\chi^*)}
\label{eqWkappa}
\end{equation}
for $\chi < \chi^*$ and zero otherwise. Here $a(\chi)$ is the scale factor and $\chi^*$ is the radial comoving distance to the last-scattering surface.
This power spectrum is sensitive to $\Sigma m_\nu$, as shown in Fig.~\ref{figNeuMassFracDiff}, and does not depend on galaxy bias or detailed non-linear modeling. In practice it is reconstructed from CMB temperature and polarization maps using a four-point function \citep[e.g.,][]{Das:2011}. 

The CMB temperature and polarization angular power spectra, $\{ C^{TT}_\ell, C^{TE}_\ell, C^{EE}_\ell, C^{BB}_\ell \}$, are also modified by lensing, which smears the acoustic peaks by adding variance to the apparent scale of a mode, converts $E$-mode polarization into $B$-mode polarization, and adds small-scale power in $T$, $E$ and $B$ \citep[e.g.,][]{Lewis:2008}. The approximate effect of massive neutrinos is shown in Fig.~\ref{figNeuMassFracDiff} for the E-mode polarization, where we artificially amplify the effects of neutrinos on the CMB lensing, rather than the primary CMB, by varying the amplitude of the lensing potential. Increasing the neutrino mass has a similar effect to decreasing the lensing amplitude. Compared to the power spectra, the reconstructed convergence field contains more information on the neutrino mass \citep{Benoit-Levy:2012}.

Massive neutrinos also affect angular diameter distances $d_A(z)$ and the expansion rate $H(z)$, as their evolution differs from a pure radiation or pure matter component \citep[e.g.,][]{Pan:2015}. These can be measured using a `standard ruler' method that is relatively free of systematic uncertainties: the primordial oscillations in the photon-baryon fluid are imprinted in the galaxy distribution as Baryon Acoustic Oscillations (BAO). The comoving scale of the oscillations is fixed by the sound horizon at decoupling, $r_s$, which is not significantly affected by neutrino masses given current limits and is in the linear regime of density perturbations ($\approx 150$~Mpc). The observed spherically-averaged BAO angular scale for galaxies at redshift $z$ is sensitive to the parameter combination $r_s / d_V(z)$; $d_V$ is the {\it volume distance} \citep{Eisenstein:2005},
\begin{equation}
\label{eqDv}
d_V(z) \equiv \left[ cz(1+z)^2 d^2_A(z) H^{-1}(z) \right]^{1/3}.
\end{equation}
For fixed cold dark matter density, more massive neutrinos increase the total late-time non-relativistic matter content, which increases the volume distance, as shown in Fig~\ref{figNeuMassFracDiff}. 

\section{Improvements in the next decade}
\label{sec:lcdm}

\begin{figure}{
\includegraphics[width=84mm]{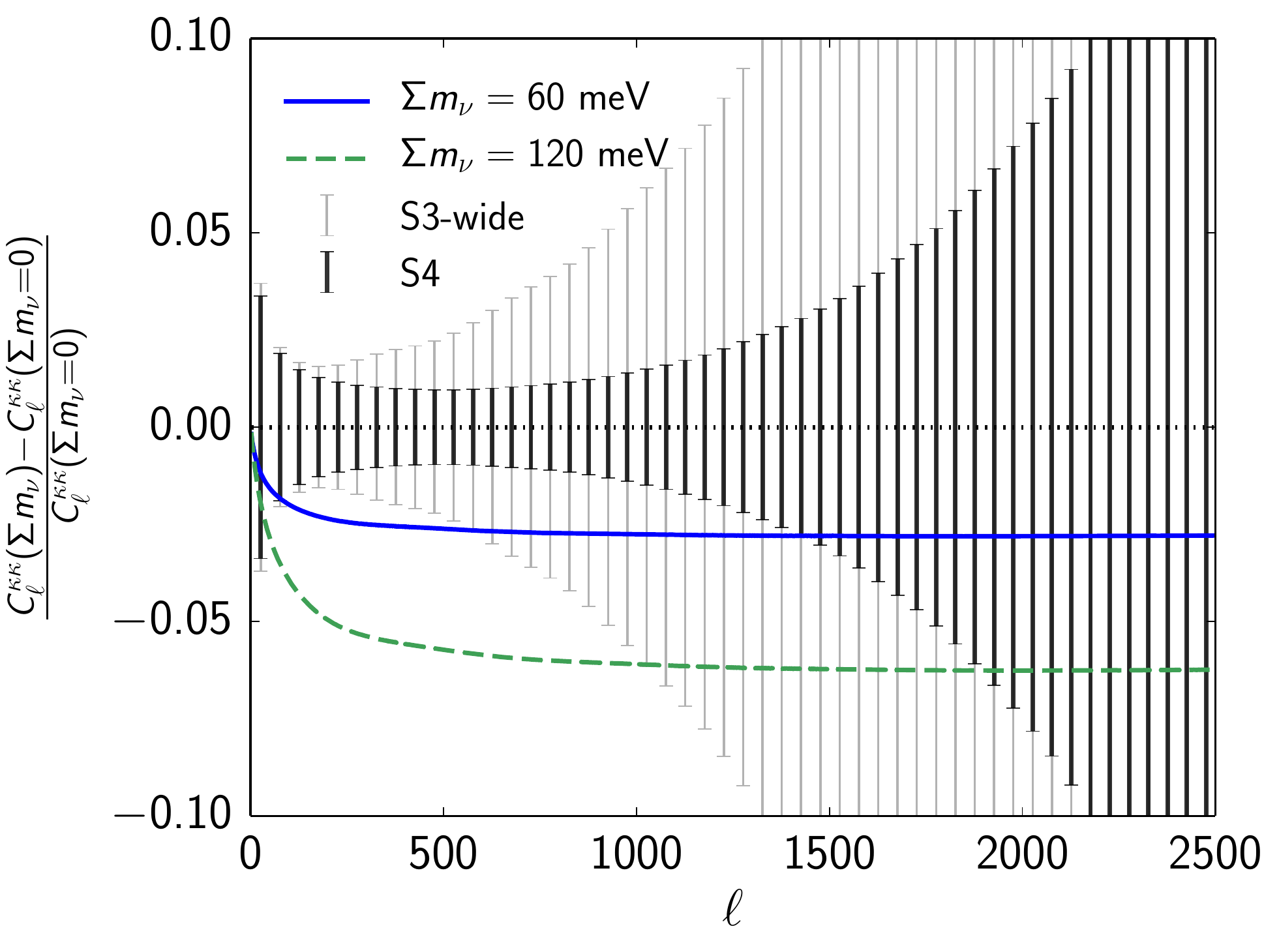}
\includegraphics[width=78mm]{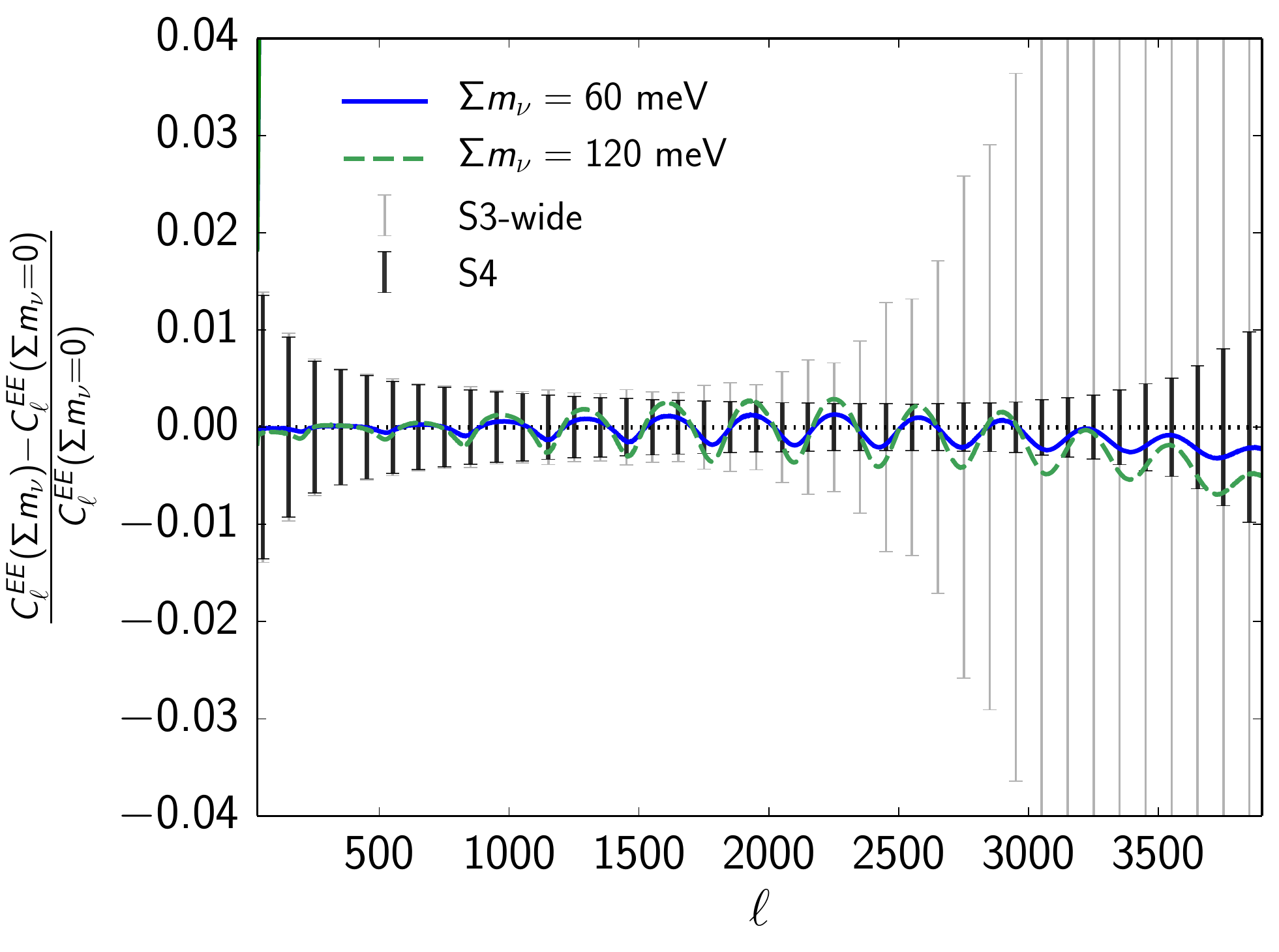}
\includegraphics[width=80mm]{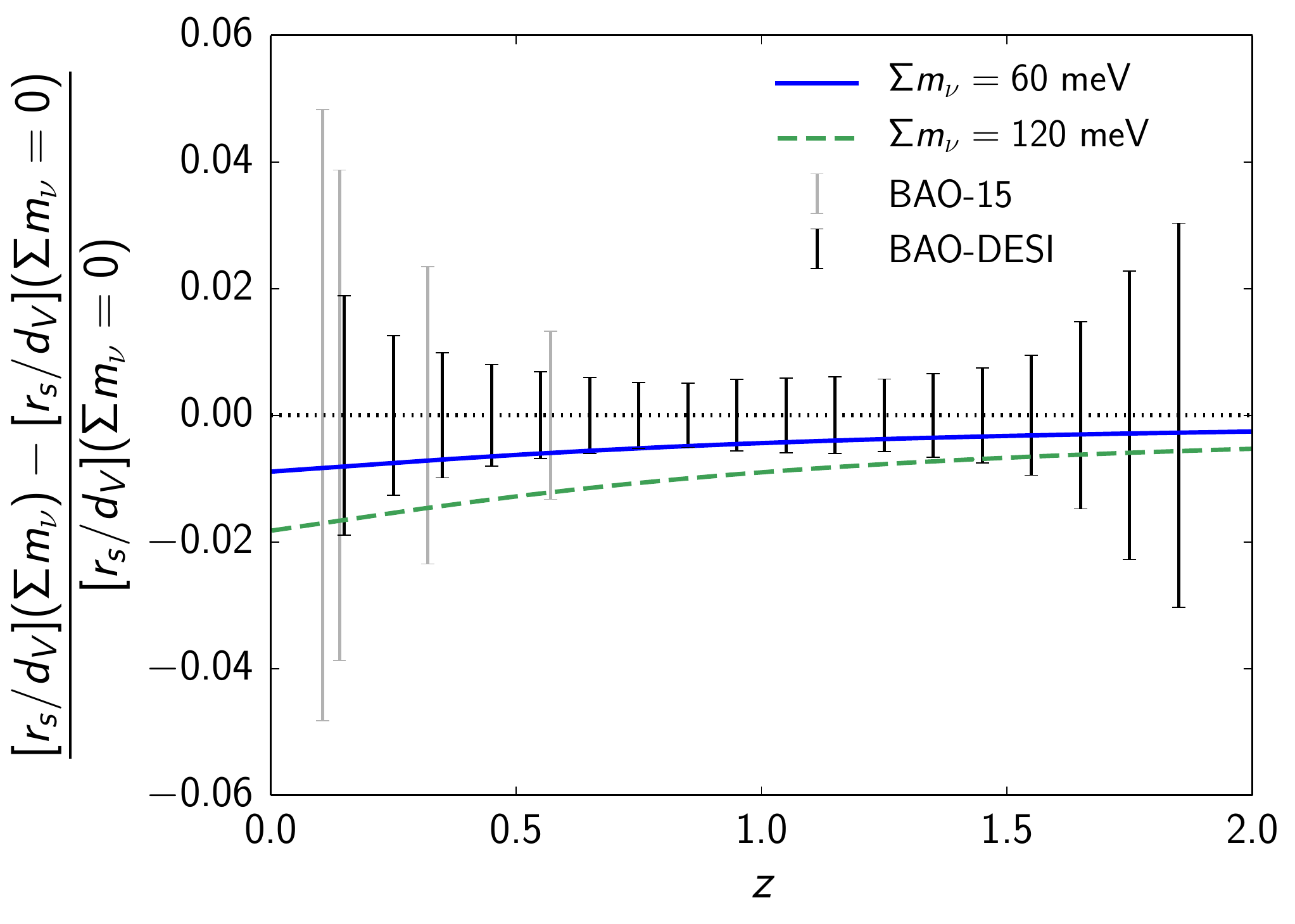}
\caption{{\sl Top and middle}: Fractional change in the convergence $\kappa$ (top) and $E$-mode (middle) power spectrum with neutrino mass, for fixed $\Omega_{\rm c}h^2 + \Omega_\nu h^2$, with expected uncertainties for S3-wide and S4 CMB data. 
A higher neutrino mass has less lensing, decreasing the E-mode peak smoothing. 
{\sl Bottom}: Fractional change in distance ratio $r_s/d_V$, with uncertainties from current (BAO-15 \cite{Anderson:2014}) and forecast (DESI, \cite{Font-Ribera:2014}) BAO data. Here $\Omega_{\rm c}h^2$ is fixed.
}
\label{fig:data}}
\end{figure}

We consider how upcoming and planned CMB and BAO experiments will improve the current limits on the sum of the neutrino masses, building on previous analyses \citep{Hall:2012,Abazajian:2013,Font-Ribera:2014, Wu:2014,Pan:2015}. 

\subsection{New data}

The current state-of-the-art for the CMB is the {\sl Planck} and {\sl WMAP} satellite data, including the first analysis of the full-mission {\sl Planck} data \citep{Planck2015XI, Planck2015XIII}.  Improved small-scale CMB measurements are currently being made by the `Stage 2' ground-based experiments: ACTPol, SPTPol and POLARBEAR \citep{Naess:2014, POLARBEAR:2014, Keisler:2015}. These will soon be upgraded to `Stage 3' (hereafter, S3) with new detectors and sensitivity to multiple frequencies. Here we consider an S3 `wide' experiment that maps 40\% of the sky, and a `deep' experiment that maps 6\% of the sky. The `wide' experiment is similar to AdvACT specifications \citep{Calabrese:2014} and `deep' to SPT-3G \citep{Benson:2014}. These experiments are expected to take data during 2016-19, and the specifications we adopt are given in Table \ref{tabCMB}. We also anticipate data from S3 experiments targeting larger angular scales (e.g., CLASS), but do not consider these specifically. In addition, we can expect a complete analysis of the {\sl Planck} polarization data, optimistically including reliable large-scale polarization data.

\begin{table}[h]
\begin{center}
\begin{tabular}{ccccc}
\hline \hline
Experiment & $f_{\rm sky}$ & Beam & $\Delta T$ &  $\Delta P$ \\ 
&  & (arcmin) & ($\mu$K-arcmin) &  ($\mu$K-arcmin) \\ 
\hline  
S3-wide & $0.4$ & 1.4 & 8.0 & 11.3  \\
S3-deep & $0.06$ &  1  & 4.0 & 5.7    \\
S4      & $0.4$ &  3 & 1   & 1.4     \\
CV-low  & $0.4$ &  60 & 1   & 1.4     \\
\hline
\end{tabular}
\end{center}
\caption{Upcoming (S3) and proposed (S4, CV-low) CMB experiments that will add to {\sl Planck}. The {\sl Planck} specifications we use are in the Appendix.}
\label{tabCMB}
\end{table}

Beyond S3, a `Stage-4' (S4) experiment - or set of experiments - is being developed by the CMB community that may cover at least half the sky to typical noise levels of $1$ $\mu$K-arcmin \citep{Abazajian:2013}. There are also proposed space-based experiments including LiteBIRD and PIXIE \citep{Matsumura:2014,kogut/etal:2011}, and we approximate their role with a cosmic variance-limited large-scale polarization measurement (`CV-low') that could supplement S3 ground-based data.

On the BAO front, current state-of-the-art measurements come from the BOSS `LowZ' and `C-MASS' galaxy samples at $z=0.32$ and $z=0.57$ \citep{Anderson:2014}. These are supplemented by data from the Six Degree Field Galaxy Redshift Survey at $z=0.11$ \citep{Beutler:2011} and the SDSS MGS sample at $z=0.15$ \cite{Ross:2015}.  Improved measurements are being made by the eBOSS survey which will survey a deeper sample \citep{eboss}. A significant advance should be made with the DESI spectroscopic survey, due to begin in 2018, which is expected to measure the BAO distance ratio from redshifts $0.15<z<1.85$ in bins of width $\Delta z = 0.1$ to percent-level precision \citep{Levi:2013,Font-Ribera:2014}. 

\subsection{Forecasting methods}
\label{ss:Forecasting}

We use a Fisher-matrix forecasting method to predict the neutrino mass uncertainties. For a model defined by parameters $\boldsymbol{\theta}$ the expected Fisher matrix is 
\begin{equation}
F_{ij}(\boldsymbol{\theta}) = \left\langle -\frac{\partial^2\ln p(\boldsymbol{\theta}|d)}{\partial\theta_i \partial\theta_j}\right\rangle,
\label{eqFisher}
\end{equation}
where $p(\boldsymbol{\theta} | \mathbf{d})$ is the posterior distribution for $\boldsymbol{\theta}$ given data $\mathbf{d}$. The forecast parameter covariance is then given by the inverse of the Fisher matrix, $\mathbb{C}_{ij}=(F^{-1})_{ij} $. Here our data are the lensed TT, TE, and EE CMB power spectra, reconstructed CMB convergence power spectrum $\kappa\kappa$, and BAO distance ratio measurements $r_s/d_V(z)$. Our parameters are the standard six $\Lambda$CDM parameters, plus the neutrino mass sum, as well as possible extension parameters including curvature and dark energy. Our methods are summarized in the Appendix, including choices made about the fiducial model, choice of parameter basis, and step-size for calculating derivatives. We also describe validation of our numerical code. 

We use the {\it lensed} CMB power spectra and the convergence power spectrum as our CMB observables, which differs from the approach in \citep{Hall:2012, Abazajian:2013, Font-Ribera:2014, Wu:2014}, but more closely follows the `real' data analysis: the CMB sky we see is lensed, and it is a difficult inverse problem to infer the unlensed sky \citep[e.g.,][]{Anderes:2015}. Using unlensed spectra in forecasts removes information contained in the lensed temperature and polarization fields. However, it is challenging to construct the full covariance matrix for the lensed power spectra and convergence power spectrum: the $T$, $Q$ and $U$ fields are all lensed by the same lensing potential, which correlates the power spectra and adds additional non-Gaussian covariance. This is explored in detail in \citep{Benoit-Levy:2012,Schmittfull:2013}. In this analysis we make the approximation of discarding BB information, and assuming Gaussian uncorrelated errors in TT, TE, EE, and $\kappa\kappa$. This is likely a good approximation for S3 data, but could under-estimate certain parameter errors for S4-type data by up to $\approx 20\%$ \citep{Benoit-Levy:2012}. We use {\sc camb} for evaluation of all relevant CMB power spectra \citep{Lewis:1999}.

For the noise levels of {\sl Planck}, we consider two cases, `{\sl Planck}-2015' (P15) that produces cosmological constraints which closely match the published results \citep{Planck2015XIII}, and `{\sl Planck}-pol' which includes TE and EE data coming from the polarization measurements of the High-Frequency Instrument (HFI), including large-scales. The specifications are given in the Appendix. For {\sl Planck}-pol, noise levels are approximated by taking temperature sensitivities from P15 and assuming the per-channel noise scalings from temperature to polarization in the {\sl Planck} Blue Book \citep{PlanckBlueBook}. This is likely to be over-optimistic at the largest scales.

For the CMB power spectra, we set a maximum multipole for the recoverable information: $\ell^T_{\rm max} = 3000$, $\ell^P_{\rm max} = 4000$ and $\ell^\kappa_{\rm max} = 3000$ for the future S3 and S4 experiments.  Smaller scales are likely hard to extract due to extragalactic foreground contamination. We assume white noise, and do not include additional foreground uncertainty beyond the multipole cuts outlined above, although the expected S3 white noise level includes some foreground inflation \citep{Calabrese:2014}. We also set a minimum multipole of $\ell_{\rm min}=50$ for S3 due to the challenge of recovering large-scales from the ground, and consider two options for S4: $\ell_{\rm min}=50$ and $\ell_{\rm min}=5$. For S3 and S4 we include {\sl Planck} data for $2<\ell <\ell_{\rm min}$, and our nominal analysis uses `{\sl Planck}-pol' unless stated otherwise. We consider the importance of the large-scale polarization measurements in \S\ref{ss:LowEllP}.
 
We use the quadratic-estimator formalism of \citet{Hu/Okamoto:2002} to calculate the CMB convergence noise spectrum $N_\ell^{\kappa \kappa}$. This uses the coupling of otherwise uncorrelated modes in temperature and polarization to reconstruct the lensing potential. Iterative delensing procedures are able to reduce the effective noise level of the estimated lensing field, particularly for the low-noise ($\Delta P \lesssim 4\mu$K-arcmin) future experiments considered here \citep{Smith:2012}. We consider the impact of this process in \S\ref{ssResults}.

For the BAO measurements we use the published uncertainties on the distance ratio $r_s/d_V$ for the current BAO data described in \S\ref{sec:lcdm} \cite{Beutler:2011,Anderson:2014,Ross:2015}, labelled as `BAO-15'. For DESI we use the forecast uncertainties on $d_A(z)$ and $H(z)$ given in \citet{Font-Ribera:2014} to estimate the expected $r_s/d_V$ uncertainties, summarized in the Appendix. We do not use broadband shape information in the galaxy power spectra.

\subsection{Expected constraints}
\label{ssResults}

\begin{figure}
\centering
\includegraphics[width=90mm]{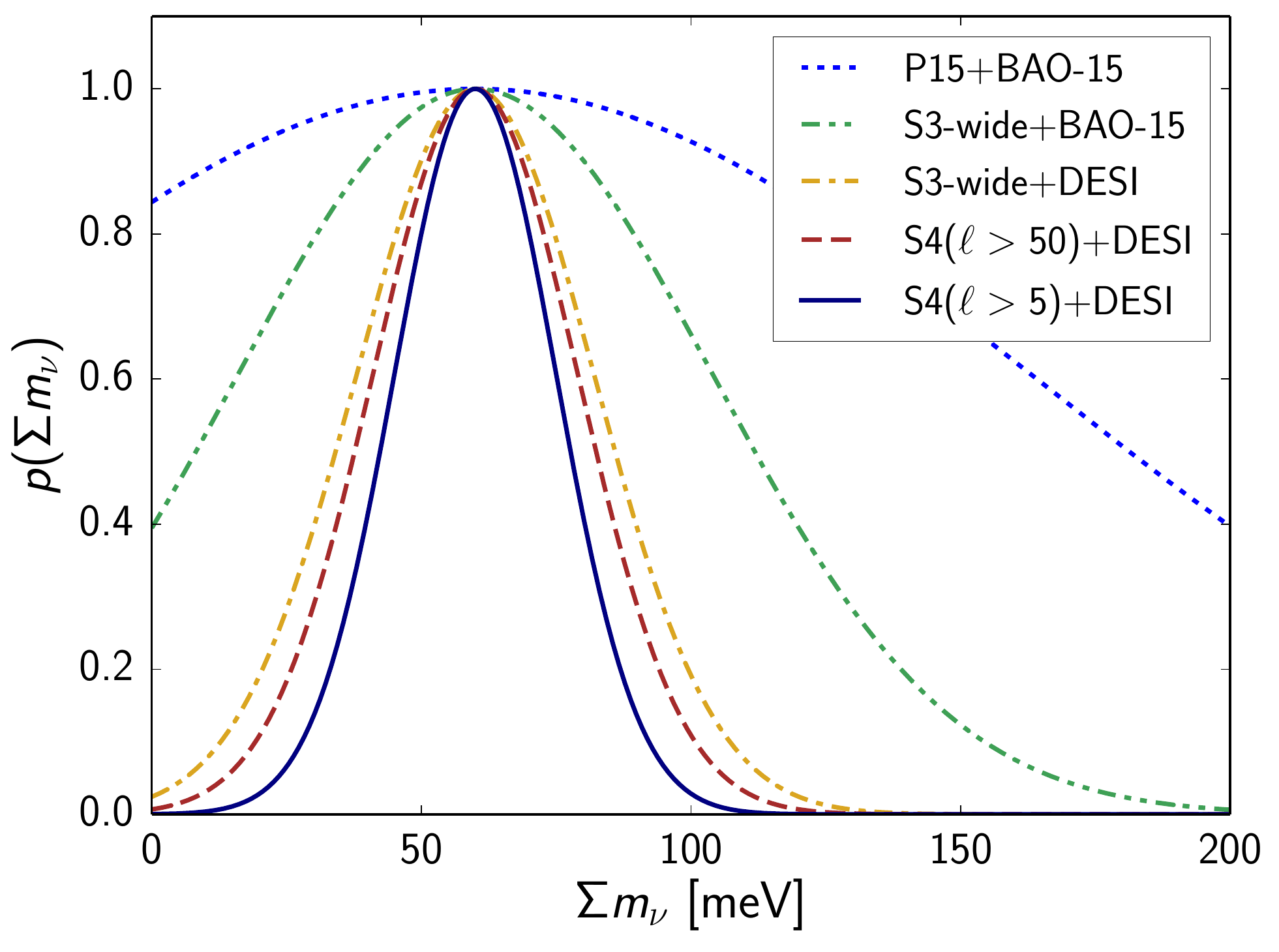}
\caption{Forecast marginal posterior constraints on the sum of the neutrino masses $\Sigma m_\nu$ within a $\Lambda$CDM+$\Sigma m_\nu$ model, assuming Gaussian error distributions. The current uncertainties (P15+BAO-15) are expected to improve rapidly, with S3 CMB data and DESI BAO data expected by $\sim$2020.
}
\label{fig:mnuOneD}
\end{figure}
\begin{figure}
\centering
\includegraphics[width=85mm]{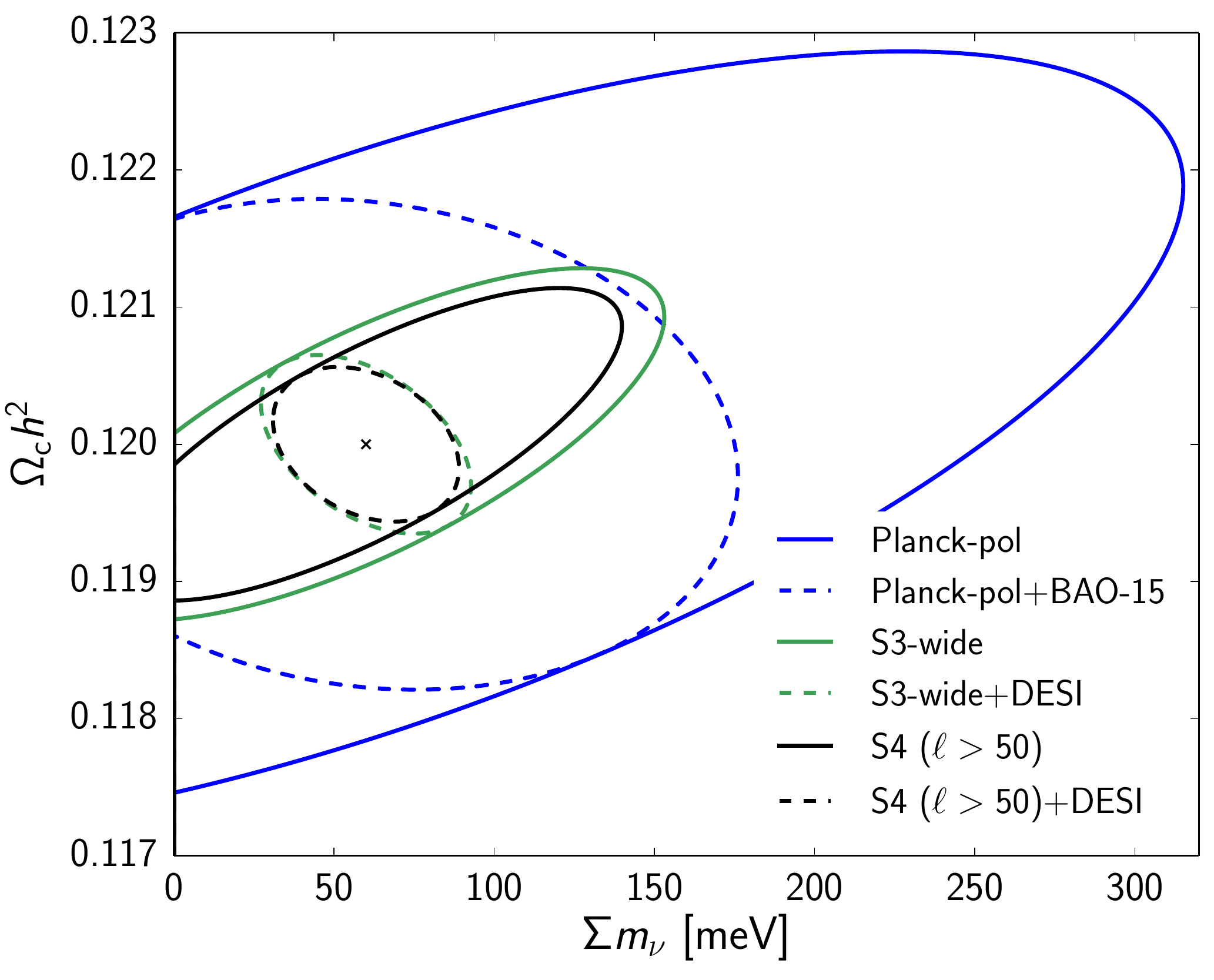}
\caption{Expected joint constraint (68\% CL) on the neutrino mass sum $\Sigma m_\nu$ and physical cold dark matter density $\Omega_{\rm c}h^2$ within a $\Lambda$CDM+$\Sigma m_\nu$ model. The BAO constraint, sensitive to the total late-time cold dark matter density, is almost orthogonal to the CMB lensing constraint, breaking the degeneracy.}
\label{fig:omch2}
\end{figure}
We find forecast marginal 1$\sigma$ uncertainties on the sum of the neutrino masses $\Sigma m_\nu$ of
\begin{equation}
\label{eqMainResult}
 \frac{\sigma(\mnum)}{\rm meV} = \begin{cases}
103	& \text{(P15 + BAO-15)}\\
44	& \text{(S3-wide  + BAO-15)}\\
22	& \text{(S3-wide  + DESI)} \\
19	& \text{(S4 ($\ell>50$) + DESI)} \\
15      & \text{(S4 ($\ell>5$) + DESI)} \\
\end{cases}
\end{equation}
where S3 and S4 include {\sl Planck}-pol at large-scales. If we replace S3-wide with the S3-deep survey we find $\sigma(\Sigma m_\nu)=53$~meV combined with BAO-15, and $25$~meV with DESI. The expected constraints are summarized in Fig.~\ref{fig:mnuOneD}, and are consistent with findings in \citet{Wu:2014}. These forecasts imply that if the neutrino hierarchy is inverted, with mass sum $>100$~meV, we may have $> 2\sigma$ evidence for non-zero mass in the next few years from S3 data, and an almost $5\sigma$ detection in $\approx$5 years with DESI. If instead it is the minimal mass of $60$~meV, a $2-3\sigma$ indirect measurement should be reachable in five years, with stronger evidence from the subsequent experiments. 

As illustrated in Fig.~\ref{fig:omch2}, there is a strong positive correlation between the neutrino mass and cold dark matter density in the CMB observables. This arises predominantly from the competing influence of these parameters on the lensing signal. Increased neutrino mass suppresses small-scale power, while increasing cold-dark matter density {\it boosts} small-scale power by shifting matter-radiation equality to earlier times; this shortens the radiation-domination epoch in which sub-horizon modes of the gravitational potential decay, enhancing the small-scale amplitude of structure \cite{Namikawa:2010, Hall:2012, Pan:2014}. Conversely, BAO data constrain the {\it sum} of the CDM and massive neutrino density, since it is this combination that affects the angular diameter distance and expansion rate.

The power of the BAO and CMB data lie in their combination. BAO measurements alone cannot constrain the primordial parameters $A_s$ and $n_s$, nor the optical depth to reionisation $\tau$; the constraints lie in the 4-dimensional subspace spanned by $\Omega _ch^2, \Omega_bh^2, \Sigma m_\nu$ and $\theta_A$ (being most constraining in the directions corresponding approximately to $\Omega_mh^2$ and $\theta_A$). S4-alone could achieve a neutrino mass error of $\sigma(\Sigma m_\nu) = 53$ meV; with DESI BAO this is expected to tighten to $\sigma(\Sigma m_\nu) = 19$ meV, enough for a 3$\sigma$ detection of the minimal mass.

Comparing to previous results, the S4($\ell>5$)+DESI forecast matches previous results \citep{Abazajian:2013, Wu:2014, Pan:2015}, and the `current data' P15 + BAO-15 forecast is compatible with the published result: $\sigma(\mnum) < 0.23$ eV at 95\% confidence \citep{Planck2015XIII}, with the forecast errors on other $\Lambda$CDM parameters also matching closely. Here we chose a fiducial neutrino mass sum of 60~meV. We find a smaller uncertainty for higher neutrino mass (as in \cite{Hall:2012}), but the effect is small when BAO data are included. For S4 alone, we find $\sigma(\Sigma m_\nu) = 53$ $(41)$ meV for a fiducial mass of 60 (120) meV, but with DESI this difference reduces to $\sigma(\Sigma m_\nu) = 19$ $(18)$ meV. We describe further tests of the code in the Appendix. 

We consider the impact of using an improved, likelihood-based lensing estimator (going beyond the first-order quadratic estimator of \citet{Hu/Okamoto:2002}) to reduce the effective noise power $N_l^{\kappa \kappa}$ of the lensing reconstruction. \citet{Hirata:2003} show that $N_l^{\kappa \kappa}$ can be reduced by a factor of two for an S4-like experiment; under this modification, we find only a 3\% tightening of the neutrino mass constraint for S4+DESI. The improvement is small due to the contribution of cosmic variance (CV) to the lensing power spectrum uncertainties, and the degeneracy of neutrino mass with other $\Lambda$CDM parameters; the S4 lensing power spectrum derived from the quadratic estimator is already CV-limited out to $\ell \approx 800$.
 
As in \citet{Benoit-Levy:2012}, we find that there is useful information in the lensed power spectra. We show the relative impact on forecasted constraints for S3-wide and S4 in Table~\ref{tabLens} (including {\sl Planck}-pol at large scales but no BAO). Lensing reduces the neutrino mass uncertainty by a factor of $\approx 6$ compared to the unlensed CMB. As such, the two-point functions will provide increasingly more information as E-mode polarization measurements improve. Checking for consistency between the lensed observables will be an important systematic test for new CMB data. For example, marginalizing over the lensing effect in the 2-point functions, by introducing a variable lensing amplitude parameter $A_{\rm lens}$ \citep{Calabrese:2008}, would isolate the impact of neutrinos on the 4-point function. 

\begin{table}[h]
\begin{center}
\begin{tabular}{cccc}
\hline \hline
 & Unlensed & Lensed TT, TE, EE  & Unlensed+$\kappa \kappa$ \\
 & & (2-point only) &      (4-point only) \\
\hline  
S3, $\sigma(\Sigma m_\nu)$:	&	435 & 75  & 61  \\
S4, $\sigma(\Sigma m_\nu)$: 		&	363 & 64  & 53 \\
\hline
\end{tabular}
\end{center}
\caption{Impact of lensing on the neutrino mass constraint (in units of meV). Constraining power comes from both the lensed spectra (2-point) and the reconstructed lensing potential (4-point). Gaussian uncorrelated errors are assumed.}
\label{tabLens}
\end{table}

\section{Dependence on experimental details}
\label{sec:depExpDetails}
Since future CMB experiments are currently under development, we investigate the importance of certain experimental details on the mass constraint. 

\subsection{Importance of the reionization bump}
\label{ss:LowEllP}
The amplitude of primordial power, $A_s$, is partially degenerate with $\Sigma m_\nu$, since $A_s$ increases the amplitude of clustering at small-scales, and $\Sigma m_\nu$ decreases it. The amplitude $A_s$ is not well determined by the primordial CMB temperature anisotropy; an increased optical depth to reionization lowers the signal such that the normalization of the anisotropy measures the combination $A_s e^{-2\tau}$ \citep{PlanckXVI}. This leads to a degeneracy between $\tau$ and $\Sigma m_\nu$ which can be broken with precision measurements of the reionisation bump at multipoles $\ell <20$ in polarization \citep{Smith:2006, Font-Ribera:2014}. 

\begin{figure}
\hskip -0.2in
{\includegraphics[width=81mm]{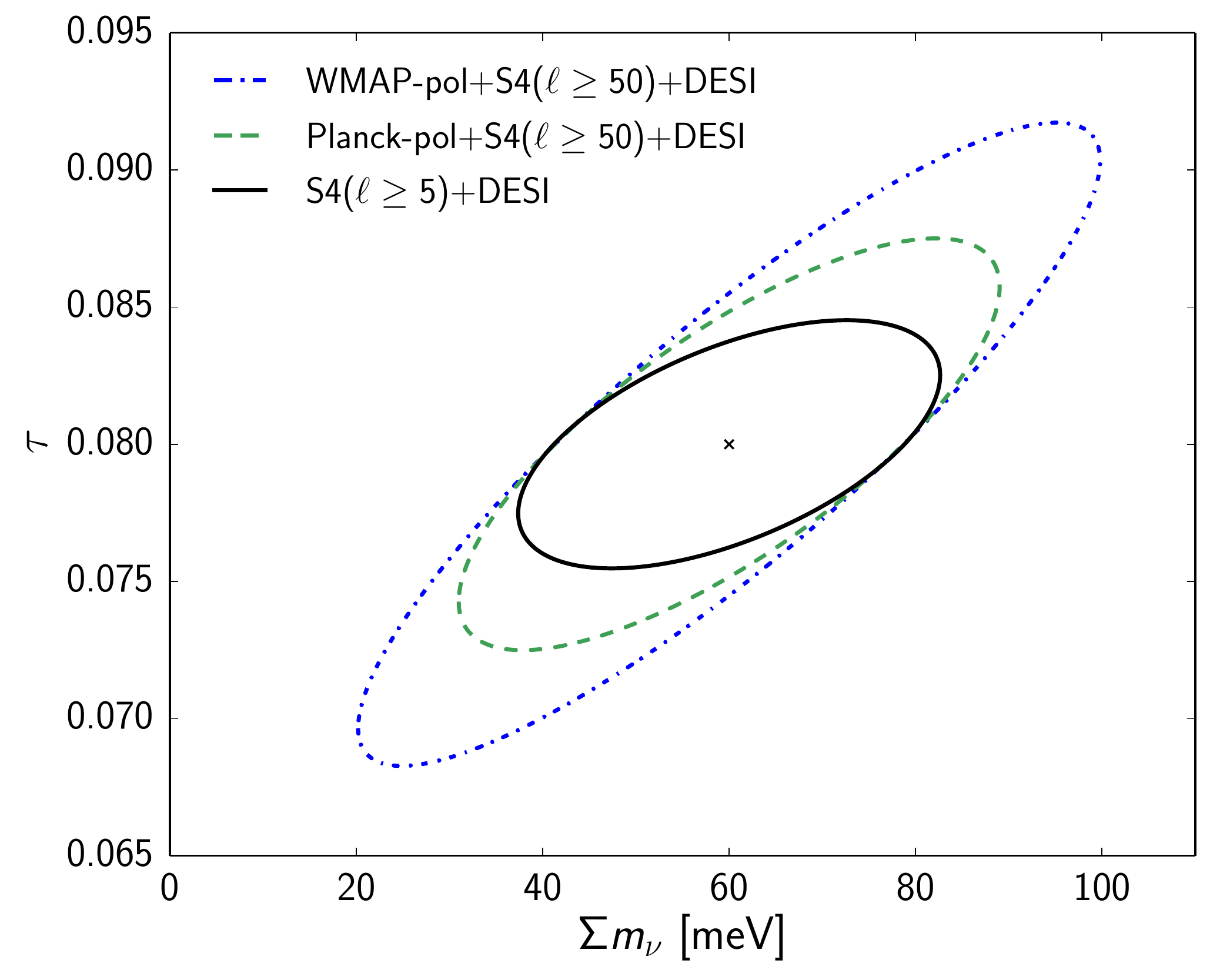}
\includegraphics[width=80mm]{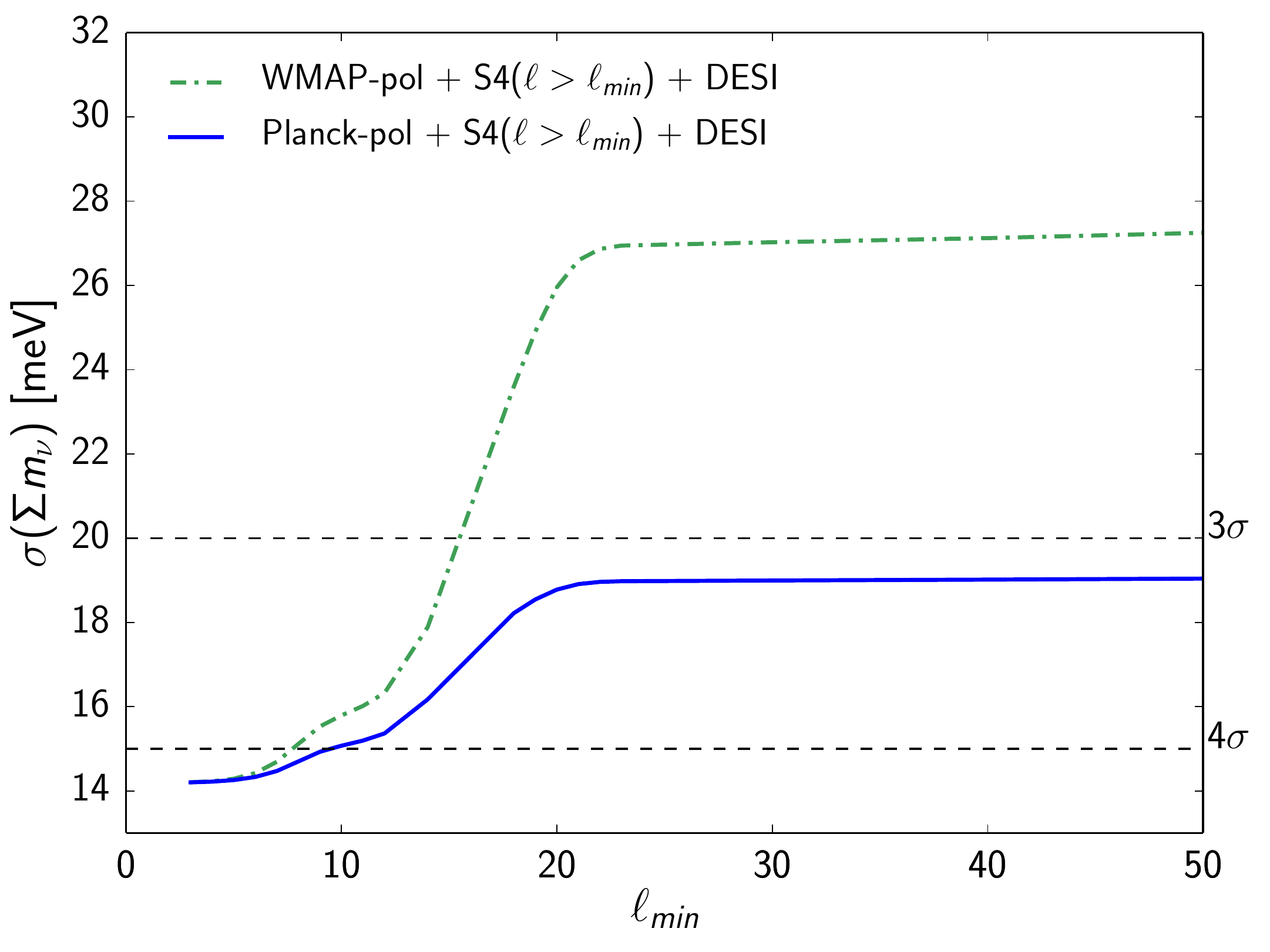}}
\caption{{\sl Top:} The neutrino mass  $\Sigma m_\nu$ is correlated with the optical depth to reionization $\tau$ (forecast 68\% CL). Current data at $\ell <20$ ({\sl WMAP}-pol) would leave a degeneracy between $\Sigma m_\nu$ and $\tau$ that could be broken with improved large-scale polarization data. {\sl Bottom:} the expected neutrino mass constraint as a function of the minimum multipole accessible to S4, indicating the benefit of reaching large scales.}
\label{figTau}
\end{figure}

Here we explore the importance of making a robust optical depth measurement, considering three cases for S4: current {\sl WMAP} measurements \citep{Hinshaw:2013, Planck2015XI}, optimistic future {\sl Planck}-pol measurements (see Appendix), and a future S4 measurement that reaches the largest scales ($\ell_{\rm min} = 5$). We find forecast constraints of 
\begin{equation}
\label{eqChangeEll}
 \frac{\sigma(\mnum)}{\rm meV} = \begin{cases}
27	& \text{(S4 $(\ell>50)$ + {\sl WMAP}-pol + DESI)}\\
19	& \text{(S4 $(\ell>50)$ + {\sl Planck}-pol \hspace{1.3mm}+ DESI)}\\
15	& \text{(S4 $(\ell>5)$ \hspace{1.8mm}+ DESI)} \\
\end{cases}
\end{equation}
with the uncertainty on $\tau$ reducing from $0.008$ to $0.005$ to $0.003$, respectively. This is compared to $0.013$ for {\sl WMAP}-pol from EE alone, i.e., improved CMB lensing data helps constrain $\tau$ even when the neutrino mass is varied. Fig.~\ref{figTau} shows the expected correlation between $\tau$ and neutrino mass. 

Fig.~\ref{figTau} also shows the impact of reducing the minimum multipole of the S4 experiment on the neutrino mass constraint, supplemented with {\sl Planck}-pol or the current {\sl WMAP}-pol at the largest scales. There is a limiting plateau for S4 at $\ell_{\rm min} >20$, and a clear improvement as the polarization is better measured at increasingly large scales. The S4($\ell>5$) + DESI limit reaches the cosmic-variance (CV) limit for CMB data\footnote{\citet{Pan:2015} found that an S4 experiment combined with a CV-limited BAO experiment could tighten the neutrino mass constraint further, to $\sigma(\Sigma m_\nu) = 11$ meV.}.

We then also consider the relative importance of making a higher sensitivity small-scale measurement, versus a new large-scale polarization measurement. We start with an S3-type $\ell>50$ experiment, and then either increase the $\ell>50$ sensitivity, or supplement it with a new `CV-low' large-scale measurement at $\ell<50$. We find forecast constraints of 
\begin{equation}
\label{eqChangeEll}
 \frac{\sigma(\mnum)}{\rm meV} = \begin{cases}
22	& \text{(S3 $(\ell>50)$ + {\sl Planck}-pol + DESI)}\\
19	& \text{(S4 $(\ell>50)$ + {\sl Planck}-pol \hspace{1.3mm}+ DESI)}\\
17	& \text{(S3 $(\ell>50)$ + CV-low + DESI)} \\
\end{cases}
\end{equation}
This indicates that a cosmic-variance-limited measurement of optical depth could be more valuable than more sensitive small-scale data, especially given that {\sl Planck}-pol large-scale polarization data is itself not yet demonstrated to be free of systematic errors.

We note that 21cm experiments, which map the brightness temperature of neutral hydrogen as a function of redshift, will probe the epoch of reionization \citep{Koopmans:2015}; combination of this information with CMB+BAO would break the $\Sigma m_\nu$-$\tau$ degeneracy and improve the neutrino mass constraint (investigated in \citet{Liu:2015}). 

\subsection{Importance of sensitivity and angular range}

\begin{figure}
\includegraphics[width=80mm]{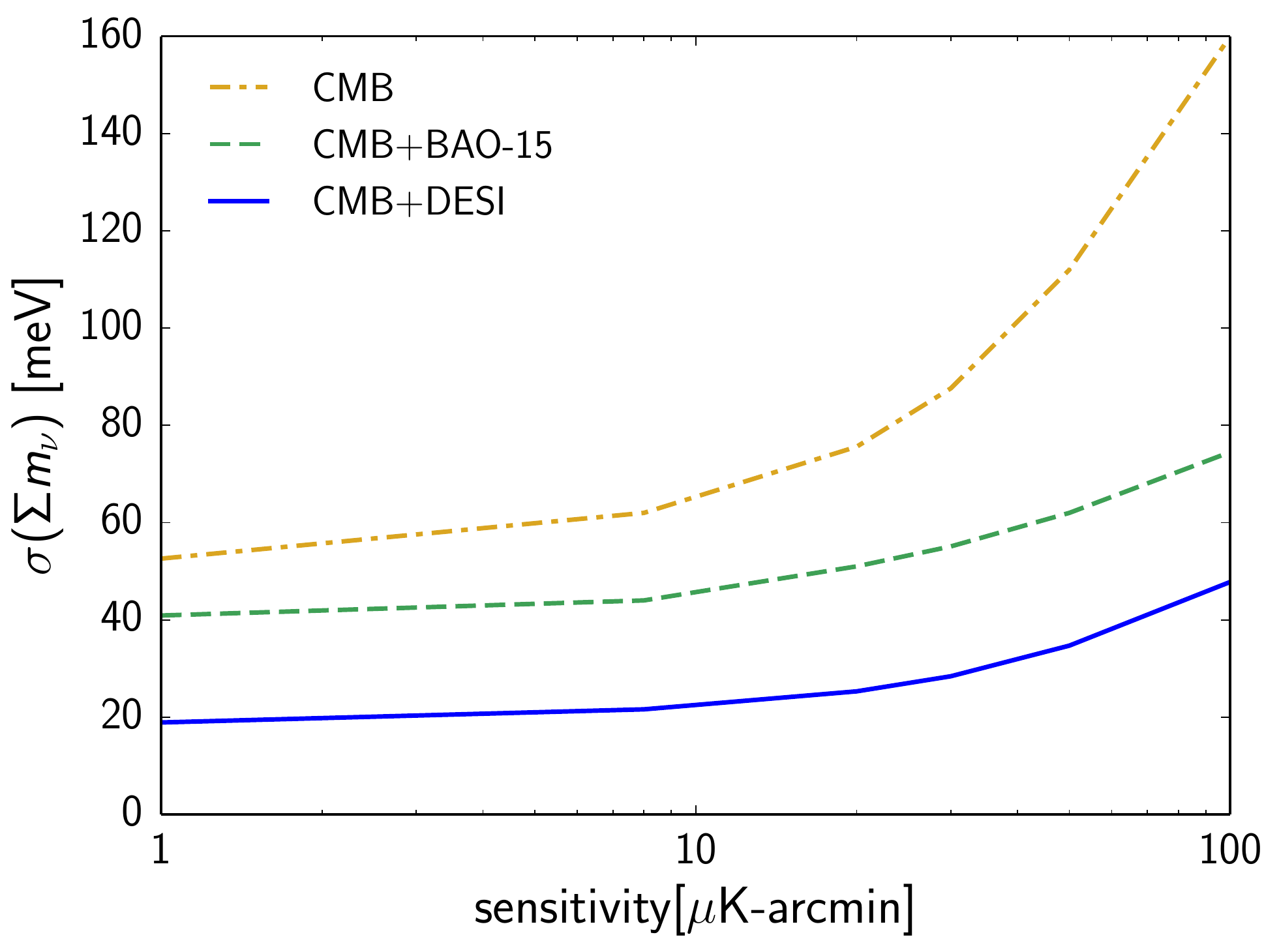}
\caption{
The dependence of the neutrino mass constraint $\sigma(\Sigma m_\nu)$ on CMB map sensitivity, for a 3 arcmin resolution experiment covering 40\% of sky.}
\label{figDelta}
\end{figure}

For the particular case of an $\ell>50$ experiment covering 40\% of the sky at 3 arcmin resolution, combined with {\sl Planck}-pol, we vary the white-noise sensitivity. The forecast neutrino mass limits are shown in Fig.~\ref{figDelta} for CMB-only, CMB+BAO-15, and CMB+BAO-DESI. There is clearly an improvement as the noise is reduced, and a significant gain is expected over current {\sl Planck} measurements, but below white noise levels of $\approx$ 10 $\mu$K-arcmin there does not appear to be a substantial gain (as also seen in \citet{Wu:2014}).

It is not yet certain whether this $\approx 10$ $\mu$K-arcmin noise level, over half the sky, will be achieved in practice from the upcoming S3 CMB experiments, or whether the lensing reconstruction will achieve the expected noise levels. Atmospheric, ground and foreground emission are typical contaminants that would increase the effective noise in the maps and in the lensing reconstruction. New data from the current S2 experiments will help clarify the impact of non-white-noise on the lensing noise performance. 

Here we have continued to restrict our analysis to `clean' scales at $\ell <3000$ in TT, TE and $\ell <4000$ in EE. A foreground-free S3 or S4 experiment would contain information from $\ell > 3000$; the $TT$ spectrum is signal-dominated to small scales ($\ell \approx 4800$ for S4) where there are a large number of modes. However, uncertainty about extragalactic foregrounds will likely make this information inaccessible. We find that including $\ell > 3000$ scales would tighten the neutrino mass error by $\approx 10\%$.

We restrict the modes available for reconstruction of the lensing potential to these same scales, and to $\ell > 50$, since reconstruction on the largest scales has yet to be demonstrated (there are difficulties such as mean-field subtraction for masked fields \citep[e.g.,][]{Planck2015XV}). We find that the information, quantified by the term $F^{\Sigma m_\nu}_\ell$ in the Fisher matrix, is concentrated in the multipole range $100 \lesssim \ell \lesssim 1000$ for S3 and $ 200 \lesssim \ell \lesssim 2000$ for S4. This also means that detailed non-linear modeling should not be required, since at redshift $ z \approx 2$ at the peak of the CMB lensing kernel (Eq.~\ref{eqWkappa}), the information peak corresponds to physical scales $\gtrsim 200$~Mpc.

\section{How unique is the massive neutrino signal?}
\label{sec:unique}
\begin{figure*}
{\includegraphics[width=140mm]{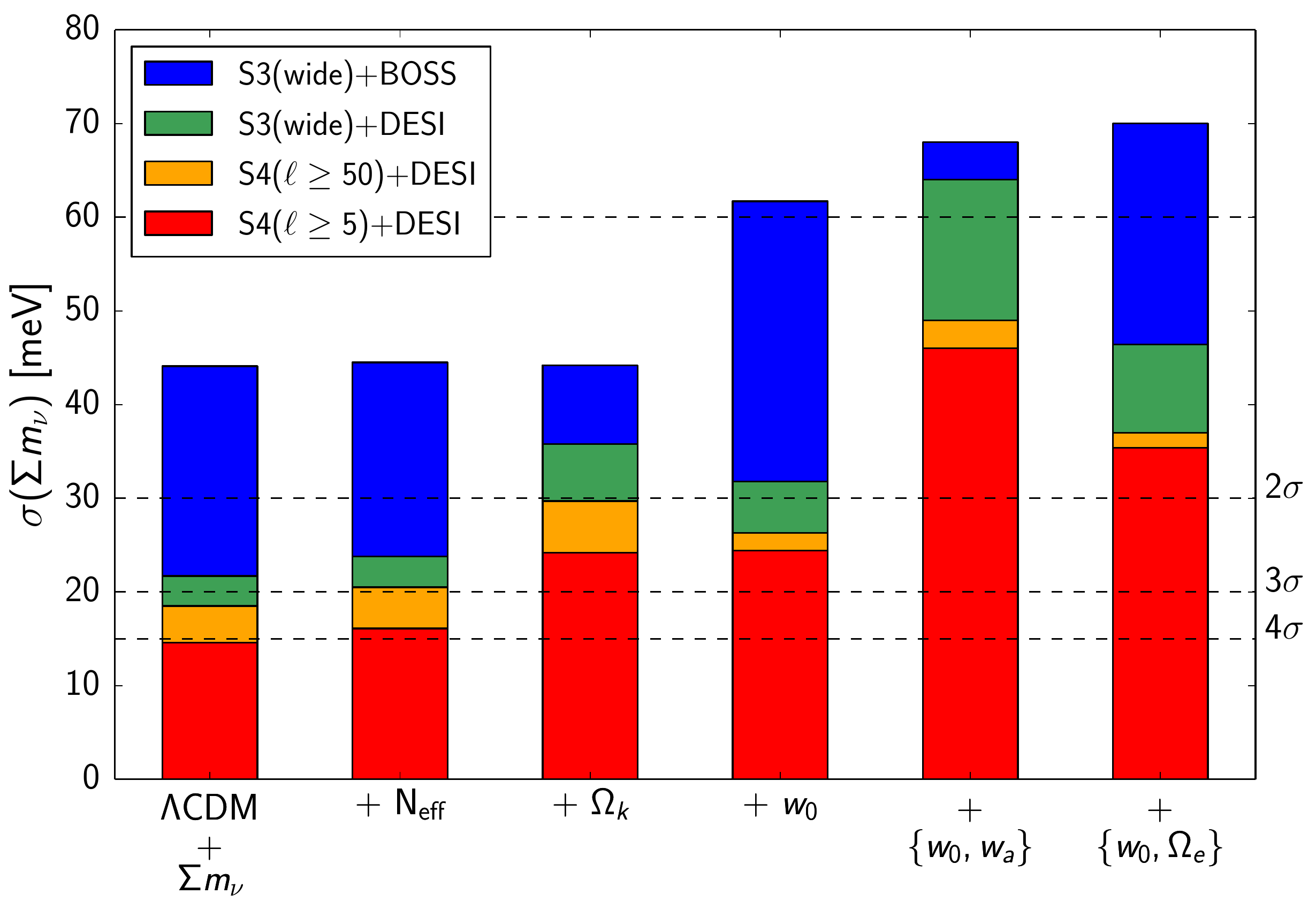}}
\caption{Neutrino mass constraints forecast for different data combinations and simple one- or two-parameter extensions to the $\Lambda$CDM+$\Sigma m_\nu$ model (all except S4($\ell \ge 5$) include {\sl Planck}-pol information at low-$\ell$). With CMB+BAO data, neutrino mass is not correlated with the number of neutrino species, $N_{\rm eff}$, but is partly correlated with spatial curvature, $\Omega_k$, and with the dark energy equation of state, $w_0$. Expected confidence levels are shown assuming the minimal total neutrino mass \mnu$=60$ meV.}
\label{figComparison}
\end{figure*}

The measurable effect of neutrinos can be partly mimicked by changes in other cosmological parameters, as explored in e.g., \citet{Hannestad:2012, Benoit-Levy:2012,Font-Ribera:2014}. Massive neutrinos affect the expansion rate and angular diameter distance, but changes in curvature, dark energy history or the Hubble parameter can compensate to keep the well-constrained acoustic peak positions and structure essentially unchanged. 

Here we consider changes in the spatial curvature, and changes in the energy density of dark energy with time. Dark energy is invoked to explain the acceleration of the universe, and may take the form of a cosmological constant with constant energy per unit proper volume, although an evolving dark energy equation of state ($w \ne -1$) is not ruled out by current observations \citep{Planck2015XIII}. We consider two parameterizations for dark energy: these are the usual Taylor expansion in the scale factor for dynamical dark energy \citep{Linder:2003};
\begin{equation}
w(a) = w_0 + w_a(1 -a),
\end{equation}
with two free parameters $w_0$ and $w_a$, and also the \citet{Doran:2006} model for early dark energy, 
\begin{equation}
\Omega_\Lambda(a) =  \frac{\Omega_\Lambda - \Omega_e \left(1- a^{-3 w_0}\right) }{\Omega_\Lambda + \Omega_m a^{3w_0}} + \Omega_e \left (1- a^{-3 w_0}\right),
\end{equation}
with parameters $\Omega_e$ and $w_0$. This has a background expansion similar to a massive neutrino for periods of the evolution of the universe. Previous work has considered neutrino mass constraints within this model \citep[e.g.,][]{Calabrese:2011,Joudaki:2012}. We use the {\sc camb} PPF module \citep{Fang:2008} and a modified version of {\sc camb} from \citet{Calabrese:2011} to compute the power spectrum within these models.

We take as our baseline the S4 $(\ell>50)$+DESI experiment. Marginalizing over simple extensions to the $\Lambda$CDM+$\Sigma m_\nu$ model, we find
\begin{equation}
\label{eqExtensions}
 \frac{\sigma(\mnum)}{\rm meV} = \begin{cases}
19	& \text{($\Lambda$CDM+$\Sigma m_\nu$)}\\
30	& \text{($\Lambda$CDM+$\Sigma m_\nu$+$\Omega_k$)}\\
27	& \text{($\Lambda$CDM+$\Sigma m_\nu$+$w_0$)}\\
46	& \text{($\Lambda$CDM+$\Sigma m_\nu$+$w_0$+$w_a$)} \\
37	& \text{($\Lambda$CDM+$\Sigma m_\nu$+$\Omega_e$+$w_0$)} \\
64	& \text{($\Lambda$CDM+$\Sigma m_\nu$+$w_0$+$w_a$+$\Omega_k$)} \\
\end{cases}
\end{equation}
We discuss these parameter degeneracies in the remainder of this section. We find that marginalizing over neutrino number $N_{\rm eff}$ (and other extension parameters that modify the primordial CMB spectrum such as a running spectral index) have a $<10$\% effect on the predicted neutrino mass uncertainties. These findings are summarized in Fig.~\ref{figComparison}, which also includes the corresponding S3 forecasts. 

At first sight, the degradation of the neutrino mass estimate in the case of varying $w_0$, $w_a$ and $\Omega_k$ simultaneously appears severe, more than tripling the error bar. However, this model has three extra parameters compared to $\Lambda$CDM, and within the Bayesian framework, one can rigorously ask whether additional parameters are required by the data, quantifying the trade-off between improving the fit against an increased complexity of the model \cite[e.g.,][for discussion in the context of cosmological data]{Trotta:2007}. This model selection approach would quantify the need for additional extension parameters, and would disfavor an over-parameterized model if it is not required by the data. In practice our challenge is likely to lie in distinguishing between different one-parameter extensions to $\Lambda$CDM: are we seeing non-zero neutrino mass, or could it be similarly well-explained by a small amount of curvature, or a small deviation from a $w=-1$ equation of state? 

\subsection{Physical degeneracies}
\label{ss:Curve}

To help understand these degeneracies, we note the Hubble parameter $H(z)$ is given by
\begin{align*}
 \left[\frac{H(z)}{H_0}\right]^2 &= \Omega_r (1 + z)^4 + \Omega_m (1 + z)^3 + \\
                      &\qquad \Omega_k (1 + z)^2  + \Omega_\Lambda(z),
\end{align*}
at times after the neutrinos become non-relativistic, where $\Omega_\Lambda(z)$ is the dark energy density, constant for $w = -1$, $\Omega_r$ is the radiation density today (e.g., photons and massless neutrinos), $\Omega_m = \Omega_c + \Omega_b + \Omega_\nu$ is the matter density today (CDM, baryons and massive neutrinos) and $\Omega_\nu h^2 = \Sigma m_\nu$ $/$ $93$ eV is the physical massive neutrino density today. The angular diameter distance is given by
\begin{equation}
\label{eqDa}
\hspace{-4mm}
d_A(z)= \frac{c}{H_0(1+z)}\begin{cases}
 \frac{1}{\sqrt{-\Omega_k}} \sin \left( \sqrt{-\Omega_k} \int_{0}^{z} \frac{H_0 dz'}{H(z')} \right) & \text{$\Omega_k < 0$}  , \\
 \int_{0}^{z} \frac{H_0 dz'}{H(z')} & \text{$\Omega_k = 0$}  , \\
 \frac{1}{\sqrt{\Omega_k}} \sinh \left( \sqrt{\Omega_k} \int_{0}^{z} \frac{H_0 dz'}{H(z')} \right) & \text{$\Omega_k > 0$}. 
\end{cases}
\end{equation}

\begin{figure}
{
\includegraphics[width=80mm]{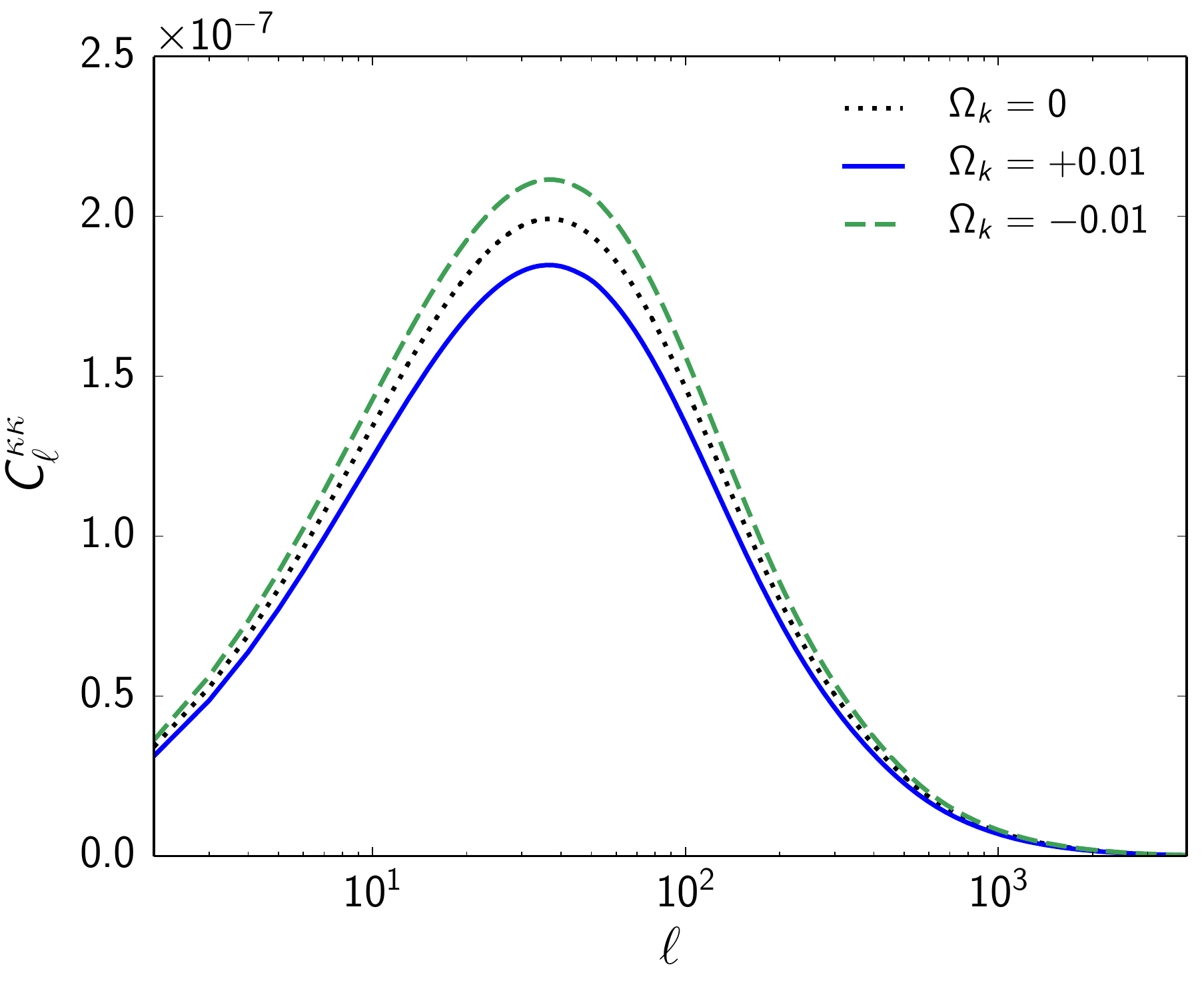}
\caption{CMB convergence power spectrum for varying $\Omega_k$, with other parameters holding the primary CMB fixed. Decreasing $\Omega_k$ requires a smaller Hubble constant $H_0$ and increased growth rate. This has a similar effect to decreasing the neutrino mass.}
\label{figClKKvarCurv}}
\end{figure}

\begin{figure}
\centering
\hskip -0.14in
\includegraphics[width=80mm]{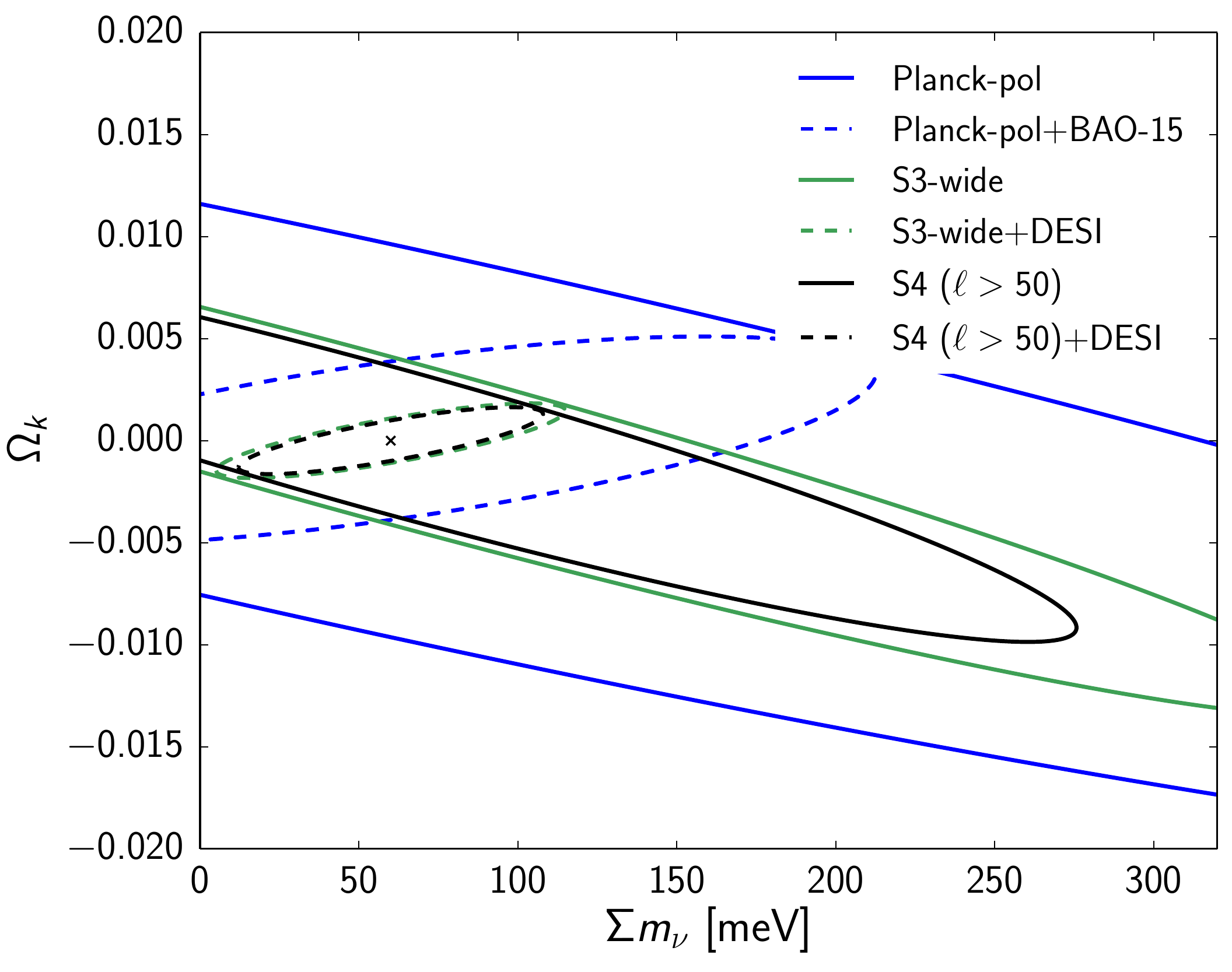}
\includegraphics[width=80mm]{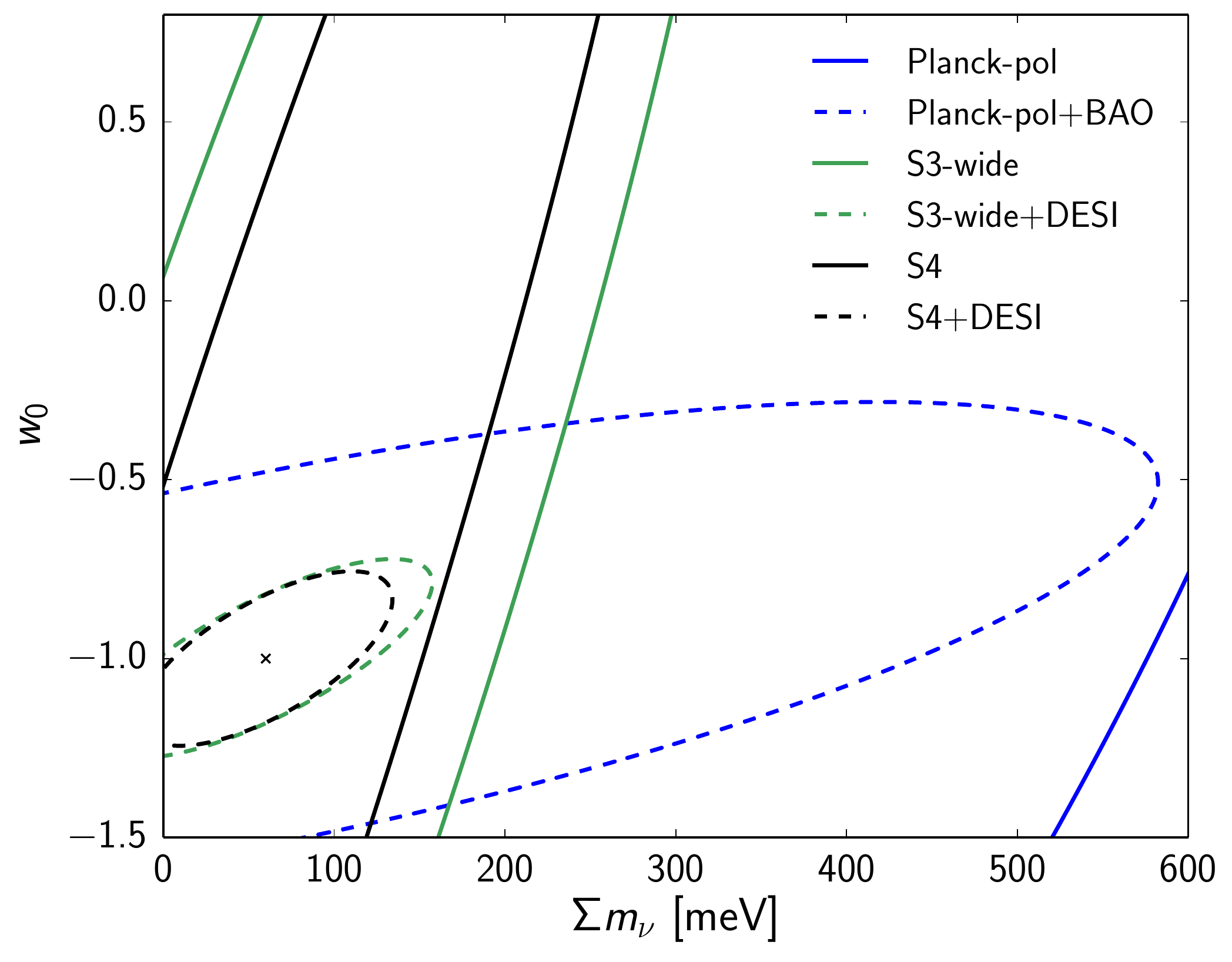}
\caption{{\sl Top:} Forecast joint constraint on the neutrino mass, $\Sigma m_\nu$, and spatial curvature, $\Omega_k$, within a $\Lambda$CDM+$\Sigma m_\nu$+$\Omega_k$ model. The BAO data breaks the anti-correlated degeneracy in the CMB data. {\sl Bottom:} Forecast constraint on $\Sigma m_\nu$ and the dark energy equation of state, $w_0$, marginalized over the $\Lambda$CDM and $w_a$ parameters. 
}
\label{fig:curve}
\end{figure}

Considering curvature, photons propagating in a non-flat universe follow curved geodesics, changing the angular diameter distance to an object of fixed proper size, at a given comoving distance, relative to a flat universe. Varying curvature shifts the angular scale of the acoustic peaks, so to remain consistent with the CMB data, the matter density and Hubble constant $H_0$ must vary to keep the peak structure unchanged. This is the well-known {\it geometric degeneracy} \citep{bond/efstathiou:1997}. 

This degeneracy is partially broken by CMB lensing measurements \cite[e.g.,][]{Sherwin:2011}, which are sensitive to the growth of structure in the late-time universe and therefore to the matter density $\Omega_m$ and dark energy density $\Omega_\Lambda$. For a fixed CMB acoustic peak scale, the effect of decreasing the curvature parameter $\Omega_k$, moving to a more closed universe, is to decrease the Hubble constant and increase $\Omega_m$. This enhances the amplitude of the CMB lensing power spectrum, as illustrated in Fig.~\ref{figClKKvarCurv}, which can be compensated by increasing $\Sigma m_\nu$. This leads to an anti-correlation between $\Sigma m_\nu$ and $\Omega_k$ when using CMB measurements alone, as shown in Fig.~\ref{fig:curve}. 

The BAO constraint in the $\Sigma m_\nu$-$\Omega_k$ plane is approximately orthogonal to the CMB-only constraint, because decreasing $\Omega_k$ decreases the volume distance to a given redshift (Eq.~\ref{eqDv}). This can be compensated by a smaller matter density, lowering the neutrino mass.  These data are powerful in combination, and for S3+BOSS the neutrino mass constraint is independent of curvature. However, the neutrino mass constraint from S3+DESI or S4+DESI is expected to degrade by $\approx 50\%$ when allowing for curvature, as illustrated in Fig.~\ref{figComparison}.

Similar arguments apply to dark energy, which modifies the background evolution according to its equation of state, but does not contribute to clustering. 
Increasing the dark energy equation of state parameter $w_0$ leads to an increased expansion rate, shifting the angular scale of the acoustic peaks, so to remain consistent with the CMB data the Hubble constant $H_0$ decreases to keep the peak structure unchanged. Similar to the curved model, this increases the clustering, which can be compensated with larger neutrino masses. This gives the positive correlation between $w_0$ and $\Sigma m_\nu$ in CMB data, illustrated in Fig.~\ref{fig:curve} and reported in e.g., \citep{Hannestad:2006,Benoit-Levy:2012}. The BAO degeneracy is also positively correlated however, so the neutrino mass uncertainty is inflated more than when allowing for curvature. Increasing $\Sigma m_\nu$ increases the contribution of neutrinos to $\Omega_m$, requiring a smaller $\Omega_\Lambda$ (in a flat universe); the volume distance to a given redshift, and hence the BAO peak position, can then be preserved by increasing $w_0$. The early dark energy density parameter $\Omega_e$, is anti-correlated with neutrino mass, due to their similar effects on the background expansion \citep{Calabrese:2011,Joudaki:2012}.
\footnote{Here we adopted a fiducial $\Omega_e = 0.007$, allowed by current data \citep{Planck2015XIV}. A smaller $\Omega_e$ would improve the neutrino mass constraint, as the parameters are anti-correlated and $\Omega_e$ cannot be negative.}

We find that the neutrino mass constraints from S3+DESI or S4+DESI are degraded by more than a factor of two when allowing for a non-minimal ($w\ne-1$) dark energy equation of state in the cosmological model, as shown in Fig.~\ref{figComparison}. Distinguishing a non-zero neutrino mass parameter from a universe with dark energy beyond the cosmological constant will be difficult with the CMB+BAO data combinations considered here. Allowing for freedom in both dark energy and curvature (i.e. a $\Lambda$CDM+$\Sigma m_\nu$+$\Omega_k$+$w_0$+$w_a$ model) degrades the constraint further to $\sigma(\Sigma m_\nu) = 64$ meV for S4+DESI, but would include three new parameters.

\subsection{Breaking degeneracies with complementary measurements}
\label{ss:Deg}

\begin{figure}
\centering
\includegraphics[width=80mm]{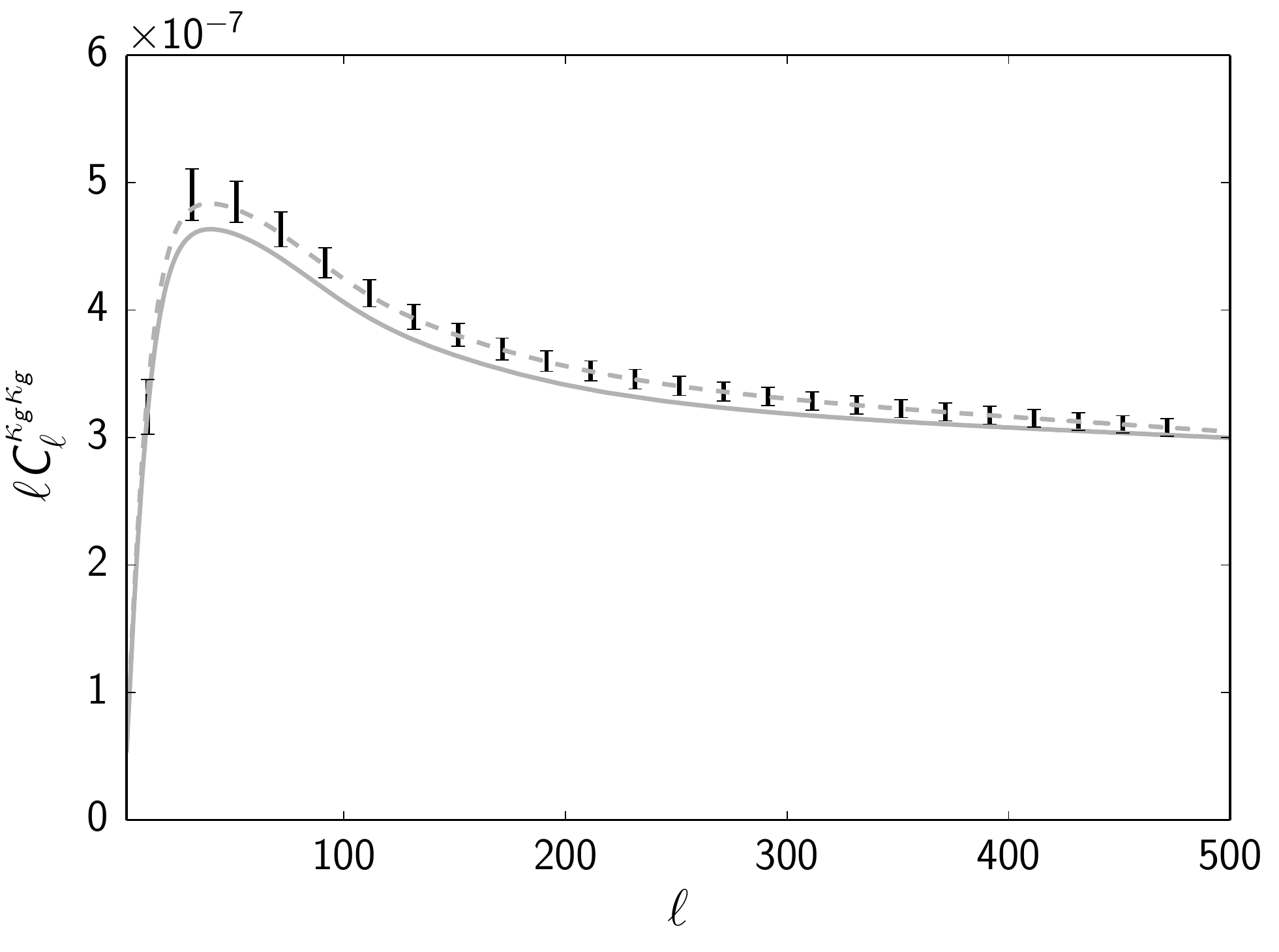}	
\caption{Galaxy lensing power spectra for two models degenerate in CMB spectra and BAO distance ratios. They could be distinguished using a future galaxy weak lensing survey ($f_{\rm sky} = 0.5$, $n_{\rm gal} = 10$ arcmin$^{-2}$, and source redshift distribution $n(z)$ as for the LSST `gold sample' \cite{LSSTscienceBookV2}). }
\label{figWL}
\end{figure}

We have focused so far on the minimal combination of future CMB and BAO data. To break these remaining degeneracies between neutrino mass and curvature/dark energy parameters, we would turn to other large-scale structure probes such as the galaxy power spectrum, redshift-space distortions, galaxy weak lensing, the kinematic Sunyaev-Zel'dovich effect and galaxy cluster counts, which measure the growth of structure at later times, and Type Ia supernovae which better constrain the expansion rate.

We have not included these and other data in our baseline forecasts as they are, arguably, more prone to systematic uncertainties such as tracer bias and the shape of the non-linear power spectrum, which contains unknown baryonic feedback effects, source distribution uncertainties and multiplicative bias \citep{Das:2013, Mandelbaum:2015, Planck2015XIII}. However, a promising path would be to examine a suite of different complementary probes, each in combination with the CMB and BAO data, to distinguish neutrino mass from a non-flat or non-$\Lambda$ model. 

For example, in Fig.~\ref{figWL} we illustrate schematically how galaxy weak lensing can break remaining degeneracies between two example models. Here the CMB spectra and BAO distances are indistinguishable with S4+DESI data (see parameters in the Appendix).

Their galaxy weak lensing signals, which probe the growth of structure during the dark energy dominated era, are distinct, and should be distinguishable with a plausible future weak lensing survey. With a 2\% difference in $\sigma_8$, robust galaxy cluster measurements should also help discriminate between these models \citep[e.g.,][]{Carbone:2012, Cerbolini:2013}.

\section{Discussion and conclusions}
\label{sec:discussion}

We have demonstrated that an indirect detection of the sum of the neutrino masses should be possible with upcoming CMB and BAO surveys. In the next decade a $4\sigma$ detection should be reachable, within the $\Lambda$CDM+$\Sigma m_\nu$ model, even in the minimal mass ($\Sigma m_\nu = 60$ meV) scenario. We have found that this is contingent on obtaining improved large-scale polarization measurements from the CMB, which may be the hardest experimental challenge. It will also be necessary to exclude degeneracies with other plausible extensions such as curvature and dark energy. We find that allowing for these extensions degrades the expected neutrino mass constraint, and that use of other large-scale structure probes will be necessary to rule out other departures from $\Lambda$CDM. 
 
Our forecasts make a number of assumptions. We have neglected non-Gaussian terms in the CMB covariance due to correlation of the temperature, polarization, and lensing fields, which should have a small impact on imminent data but will get more important as noise levels reduce. We assume Gaussianity of the posterior distribution and rely on data-independence of the covariance matrix. These assumptions become increasingly accurate for the precision future experiments considered here, but remain an approximation. Our analysis also assumes little foreground contamination. This will be most important for the large-scale CMB signal, but is not expected to significantly degrade the smaller-scale lensing signal. 

We have assumed white noise in the new CMB data. This has not yet typically been achieved in practice from the ground, due to atmospheric and scan-synchronous emission which induces additional variance at large scales. However, polarization measurements are cleaner than temperature in this respect, and new experiments have sophisticated designs to modulate the polarization signal. Performance of current `Stage-2' experiments will help refine noise projections for the future experiments. We have also assumed ideal performance of the future DESI BAO experiment. 

In terms of theoretical scope, beyond an evolving dark energy equation of state, we have not considered other possible new physics. For instance, theoretically-motivated axion contributions to the background expansions and perturbations might offer an alternative explanation for a neutrino mass detection. The cosmological effects are not identical \cite{Marsh:2012}, but further investigation will be useful. 

Finally, neutrino mass is the next beyond-$\Lambda$CDM parameter that we know will be needed to fit data, so it is valuable that competitive constraints are expected to come from different combinations of cosmological datasets beyond the CMB/BAO considered here. Their complementarity will aid in convincingly excluding systematic effects and alternative cosmological models.

\section*{Acknowledgments}
We thank David Alonso, Pedro Ferreira and Blake Sherwin for useful discussions. Some similar forecasts were done for the Advanced ACT proposal; we thank the ACT team for valuable input. RA, JD and TL are supported by ERC grant 259505, and EC is supported by STFC. PC is grateful to the University of Oxford for its hospitality while part of this work was done, and to the \'{E}cole Normale Sup\'{e}rieure for financial support.


\appendix*
\section{Forecasting parameter constraints}
\label{sFisher}

\begin{table}
 \begin{center}
  \begin{tabular}{lcc}
   \hline \hline
Parameter         & Fiducial value & Step size \\ \hline
$\Omega_b h^ 2$   & 0.0222        & 0.0008 \\ 
$\Omega_c h^2$    & 0.1197         & 0.0030 \\ 
$10^2\theta_A$ & 1.0409        & 0.0050 \\ 
$10^9A_s$             & 2.196    & 0.1 \\ 
$n_s$             & 0.9655         & 0.010 \\ 
$\tau$            & 0.078          & 0.020 \\
$\Sigma m_\nu$ (meV)   & 60           & 20 \\ 
\hline
$\Omega_k$        & 0              & 0.01 \\
$w_0$             & -1             & 0.3 \\
$w_a$             & 0              & 0.6 \\ 
$N_\text{eff}$    & 3.046          & 0.080 \\
$\Omega_e$  & 0.007          & 0.002 \\ 
\hline
  \end{tabular}
 \end{center}
\caption{Fiducial values and step sizes for the numerical derivatives, chosen to be small enough to minimize error in the Taylor expansion of the two-sided derivative, while keeping numerical stability in the derivatives from the {\sc camb} power spectra. The fiducial value of $\Omega_e = 0.007$ is at the upper 95\% confidence level given current data \citep{Planck2015XIV}.}
\label{tabParameterInifile}
\end{table}

\begin{table*}
\begin{center}
\begin{tabular}{llllllll}
\hline 
Experiment & $f_{\rm sky}$ & $\nu$/GHz & $l_{\rm min}$ &  $l_{\rm max}$  & FWHM/arcmin & $\Delta T$/$\mu$K-arcmin &  $\Delta P$/$\mu$K-arcmin \\ 
\hline 
\hline
{\sl Planck}-2015 & $0.44$ & 30,44,70,100,143,217,353    & 2 & $2500$ &  33,23,14,10,7,5,5,5  & 145,149,137,65,43,66,200 & -,-,450,-,-,-,-   \\
{\sl Planck}-pol & " &    &  &  &    &  & -,-,450,103,81,134,406 \\
{\sl WMAP}-pol & 0.74   & 33,41,64,94  & 2 & 1000 &  41,28,21,13 & -,-,298,296 & 425,420,424,-\\

\hline
\end{tabular}
\end{center}
\caption{Specification for the {\sl Planck} and {\sl WMAP} experiments used in the analysis, assuming white noise properties. We define `{\sl Planck}-2015' to reproduce the  constraints from {\sl Planck} 2015 data; for `{\sl Planck}-pol' we use the {\sl Planck} Blue Book scaling factors to convert to polarization. 
For {\sl WMAP}-pol we recover an optical depth uncertainty that matches the WMAP9 data. When combining with S3 and S4, we include {\sl Planck}-pol data across the full $f_{\rm sky} = 0.44$ at large scales ($\ell < \ell^{\rm S3/S4}_{\rm min}$) and across $f_{\rm sky} = 0.2$ in the multipole range $\ell^{\rm S3/S4}_{\rm min} < \ell < 2500$ as the useful {\sl Planck} data and S3/S4 will likely not overlap completely. Our results are insensitive to the exact choice of this non-overlapping region size. When using {\sl WMAP}-pol data we substitute it in at large scales ($\ell < \ell^{\rm S3/S4}_{\rm min}$) over $f_{\rm sky} = 0.74$.
}
\label{tabCMBExperimentalDetails}
\end{table*}

\begin{table}
\begin{center}
\begin{tabular}{lllll}
\hline 
Experiment & Redshift & $\frac{\sigma(r_s/d_V)}{(r_s/d_V)}$ & $\sigma(r_s/d_V)$  & Ref \\ 
 &  & $(\%)$ &  &  \\ 
\hline 
\hline
6dFGRS   & 0.106  & 4.83  &0.0084 & \citep{Beutler:2011}\\
SDSS MGS & 0.15 & 3.87 &0.015  & \citep{Ross:2015}\\
LOW-Z    & 0.32  & 2.35 &0.0023 & \citep{Anderson:2014}\\
C-MASS   & 0.57  & 1.33 &0.00071& \citep{Anderson:2014}\\
\hline
DESI & 0.15 & 1.89 &0.0041 & \citep{Font-Ribera:2014}\\
     & 0.25 & 1.26 &0.0017 & \\
     & 0.35 & 0.98 &0.00088 & \\
     & 0.45 & 0.80 &0.00055 & \\
     & 0.55 & 0.68 &0.00038 & \\
     & 0.65 & 0.60 &0.00028 & \\
     & 0.75 & 0.52 &0.00021 & \\
     & 0.85 & 0.51 &0.00018 & \\
     & 0.95 & 0.56 &0.00018 & \\
     & 1.05 & 0.59 &0.00017 & \\
     & 1.15 & 0.60 &0.00016 & \\
     & 1.25 & 0.57 &0.00014 & \\
     & 1.35 & 0.66 &0.00015 & \\
     & 1.45 & 0.75 &0.00016 & \\
     & 1.55 & 0.95 &0.00019 & \\
     & 1.65 & 1.48 &0.00028 & \\
     & 1.75 & 2.28 &0.00041 & \\
     & 1.85 & 3.03 &0.00052 & \\
\hline
\end{tabular}
\end{center}
\caption{Specification for current BAO-15 data (top), and forecast DESI data (bottom). We derive the expected fractional uncertainties on $r_s/d_V$ for DESI from the fractional errors on $D_A/r_s$ and $H(z)$ forecast in \citep{Font-Ribera:2014}. The absolute values correspond to a $\Lambda$CDM model with $\Sigma m_\nu=60$ meV.}
\label{tabBAO}
\end{table}

We forecast posterior parameter constraints and degeneracy directions by evaluating the relevant Fisher matrix as defined in Eq. \ref{eqFisher}. We assume uniform priors on all parameters, allowing the posterior $p(\boldsymbol{\theta})$ to be replaced with the likelihood $\mathcal{L}(\boldsymbol{\theta}) = \mathbb{P}(\mathbf{d} | \boldsymbol{\theta}, M)$ in Eq.~\ref{eqFisher}. The Fisher matrix is evaluated at {\it fiducial parameters}; the choice is unimportant under the assumption of a Gaussian posterior and data-independent covariance, although in practice they are chosen to match the current best-fit parameters of the model, as the real covariance does have a term that scales with the signal. Our choices are shown in Table \ref{tabParameterInifile}.

To compute the Fisher matrix we use as observables the CMB power spectra, and the BAO distance measurements. For the CMB, we use the lensed power spectrum between each pair of (assumed Gaussian) fields $X, Y$ from their spherical harmonic coefficients:
\begin{equation}
\label{eqEstimator}
\hat{C}^{XY}_\ell = \frac{1}{2\ell+1}\sum_{m=-\ell}^{m=\ell} x^{*}_{\ell m} y_{\ell m}.
\end{equation}
This formula omits beam smoothing effects and the subtraction of a noise term, which we account for below. The estimated power spectrum is a sum of many random variables of finite variance, and to good approximation follows a Gaussian distribution. This approximation breaks down at large scales but does not have a significant impact on expected errors. For a full-sky survey, we have 
\begin{multline}
-2\ln\mathcal{L}(\boldsymbol{\theta}) = -2\sum_\ell \ln p( \hat{C}_\ell | \boldsymbol{\theta}) \\
=  \sum_\ell  \Big[ (\hat{C}_\ell - C_\ell(\boldsymbol{\theta}) )^\top  \mathbb{C}^{-1}_\ell(\boldsymbol{\theta}) \big(\hat{C}_\ell - C_\ell(\boldsymbol{\theta})) + \ln \det(2 \pi \mathbb{C}_\ell(\boldsymbol{\theta})) \Big]
\end{multline}
where $ \hat{C}_\ell = (\hat{C}_\ell^{TT}, \hat{C}_\ell^{TE}, ...) $ contains auto- and cross-spectra and $\mathbb{C}_\ell$ is their covariance matrix. Inserting this likelihood into Eq.~\ref{eqFisher} and neglecting parameter dependence in the power spectrum covariance matrix one obtains
\begin{equation}
F_{ij} = \sum_\ell \frac{\partial C^\top_l}{\partial \theta_i} \mathbb{C}^{-1}_\ell \frac{\partial C_l}{\partial \theta_j}.
\end{equation}
From Eq.~\ref{eqEstimator}, and applying Wick's theorem, the covariance matrix for the power spectra has elements
\begin{multline}
\mathbb{C}(\hat{C}_l^{\alpha \beta}, \hat{C}_l^{\gamma \delta}) = \frac{1}{(2l+1)f_{\rm sky}} \big[ (C_l^{\alpha \gamma} + N_l^{\alpha \gamma}) (C_l^{\beta \delta} + N_l^{\beta \delta})  \\
+ (C_l^{\alpha \delta} + N_l^{\alpha \delta}) (C_l^{\beta \gamma} + N_l^{\beta \gamma}) \big],
\end{multline}
where $\alpha, \beta, \gamma, \delta \in \{T, E, B, \kappa_c\}$ and $f_{\rm sky}$ accounts for the loss of information due to partial sky coverage \citep{Hobson:1996,dePutter:2009}. 
Noise spectra are generated for each observable given input noise properties such as CMB map sensitivities. We assume additive white-noise for the CMB:
\begin{equation}
N^{\alpha \alpha}_\ell = (\Delta T)^2 \exp \left( \frac{\ell(\ell + 1) \theta^2_{\rm FWHM}}{8 \ln 2} \right)
\end{equation}
for $\alpha \in \{T, E, B\}$, where $\Delta T$ ($\Delta P$ for polarization) is the map sensitivity in $\mu$K-arcmin and $\theta_{\rm FWHM}$ is the beam width. This as an optimistic approximation: real noise-spectra from ground-based experiments have a dominant contribution from atmospheric variance at large scales (see e.g. Fig. 4 of \citet{Das:2013}). The atmosphere is weakly polarized, and hence the white-noise approximation is better in $E$ and $B$ than $T$.
The CMB lensing reconstruction noise is calculated using the \cite{Hu:2002} quadratic-estimator formalism. As described in the main text, we neglect non-Gaussian terms in the power spectrum covariance, and also neglect the BB spectrum as it does not contribute significantly to upcoming constraints and has a highly non-Gaussian covariance \citep{Benoit-Levy:2012}. 

We add information from Baryon Acoustic Oscillation (BAO) experiments by computing the BAO Fisher matrix:
\begin{equation}
F_{ij}^{\rm BAO} = \sum_{k} \frac{1}{\sigma_{f,k}^2}\frac{\partial f_k}{\partial \theta_i}\frac{\partial f_k}{\partial \theta_j}
\end{equation}
where $f_k \equiv f(z_k) = r_s/d_V(z_k)$ is the sound horizon at photon-baryon decoupling $r_s$ over the volume distance $d_V$ to the source galaxies at redshift $z_k$. These real and forecast data are reported in Table \ref{tabBAO}.

The total Fisher information matrix is given by the sum of the CMB and BAO Fisher matrices, and is inverted to forecast parameter covariances.  An alternative MCMC approach using simulated data can be taken to account for non-Gaussianity of the posterior \citep[e.g.,][]{Hall:2012}, but the Gaussian approximation is likely increasingly good as the data quality improve from {\sl Planck} through S3 to S4. 

Our forecasting code, {\sc OxFish}, has been developed for this analysis and is used to forecast parameter covariance matrices in one coherent python package. The code interfaces with the {\sc camb} code for evaluation of power spectra. We compare with real data or previous work where possible.  We construct the `Planck-2015' (P15) specification, given in Table~\ref{tabCMBExperimentalDetails}, to produce constraints which match the $\Lambda$CDM uncertainties in \citet{Planck2015XIII}, with the beam sizes and noise levels matching the detector sensitivities in \cite{Planck2015I}. 

We also forecast the neutrino mass constraint from P15+BAO, finding $\sigma(\Sigma m_\nu) = 103$ meV. Placing the peak of the posterior at the fiducial $\Sigma m_\nu = 60$ meV, this corresponds to $\Sigma m_\nu < 245$ meV at 95\% confidence, comparable to the actual result of $\Sigma m_\nu < 230$ meV \cite{Planck2015XIII}. This includes JLA SNe data and an $H_0$ prior, but these are expected to have a small impact. We also agree with \cite{Abazajian:2013, Wu:2014, Pan:2015} on the neutrino mass constraint for the S4+DESI data combination, finding $\sigma(\Sigma m_\nu) = 15$ meV if we assume that the reionization bump is measured.

We use the following parameters for the curves in Fig.~\ref{figWL}: solid curve; \{$\Omega_bh^2=0.0222$, $\Omega_ch^2=0.0120$, $\Sigma m_\nu = 3$ meV, $\tau = 0.067$, $10^2\theta_A = 1.0417$, $10^9A_s = 2.15$, $n_s = 0.9647$, $\sigma_8=0.835$, $H_0=68.0$\}, dashed curve; \{$\Omega_bh^2 = 0.0222$, $\Omega_ch^2 = 0.0119$, $\Sigma m_\nu = 117$ meV, $\tau = 0.089$, $10^2\theta_A = 1.0399$, $10^9A_s = 2.24$, $n_s = 0.9663$, $\sigma_8=0.822$, $H_0=66.6$\}.

\label{lastpage}
\clearpage
\end{document}